\documentclass[a4paper,12pt]{article}

\usepackage{amsmath}
\usepackage{amssymb}
\usepackage{graphicx}
\usepackage[OT4]{fontenc}  
\usepackage[utf8]{inputenc}
\usepackage{url}
\usepackage[square,numbers]{natbib}
\usepackage[unicode,debug,colorlinks=true,linkcolor=black,citecolor=black,urlcolor=black,pdfusetitle=true,breaklinks=true]{hyperref}
\usepackage{setspace}

\newcommand{\abs}[1]{\left| #1 \right|}
\newcommand{\norm}[1]{\left\| #1 \right\|}
\newcommand{\okra}[1]{\left( #1 \right)}
\newcommand{\mean}[1]{\mathbf{E}\, #1}
\newcommand{\ceil}[1]{\left\lceil #1 \right\rceil}
\newcommand{\floor}[1]{\left\lfloor #1 \right\rfloor}
\newcommand{\kwad}[1]{\left[ #1 \right]}
\newcommand{\klam}[1]{\left\{ #1 \right\}}

\newcommand{\boole}[1]{{\bf 1}{\klam{#1}}}

\DeclareMathOperator{\card}{\#}

\DeclareMathOperator*{\hilberg}{hilb}

\DeclareMathOperator*{\argmin}{arg\,min}

\newtheorem{definition}{Definition}
\newtheorem{corollary}{Corollary}
\newtheorem{theorem}{Theorem}
\newtheorem{lemma}{Lemma}
\newenvironment*{proof}{\begin{trivlist}\item[]
\noindent\textbf{Proof:}}{ \hfill$\Box$\par\end{trivlist}}
\newenvironment*{proof*}[1]{\begin{trivlist}\item[]
\noindent\textbf{Proof of #1:}}{ \hfill$\Box$\par\end{trivlist}}

\begin{document}

\title{Local Grammar-Based Coding Revisited}

\author{{\L}ukasz D\k{e}bowski %
  \thanks{This work was done in part during the support of the
    National Science Centre Poland Grant 2018/31/B/HS1/04018.  The
    work is an extension of a manuscript under the same title
    that contained significantly fewer results
    (\url{https://arxiv.org/abs/2209.13636v2}).  Some results of this
    work have been mentioned in the draft of the unpublished and
    unrefereed textbook ``A Short Course in Universal Coding''
    (\url{http://dx.doi.org/10.13140/RG.2.2.21948.24961}), whereas
    Lemma 1 has been mentioned in the published article
    \cite{Debowski25}.} %
  \thanks{{\L}. D\k{e}bowski is with the Institute of Computer
    Science, Polish Academy of Sciences, ul.\ Jana Kazimierza 5,
    01-248 Warszawa, Poland (e-mail: \url{ldebowsk@ipipan.waw.pl}).} %
}

\date{}

\begin{titlepage}
\maketitle

\begin{abstract}
  In the setting of minimal local grammar-based coding, the input
  string is represented as a grammar with the minimal output length
  defined via simple symbol-by-symbol encoding. This paper discusses
  four contributions to this field. First, we invoke a simple harmonic
  bound on ranked probabilities, which reminds Zipf's law and
  simplifies universality proofs for minimal local grammar-based
  codes. Second, we refine known bounds on the vocabulary size,
  showing its partial power-law equivalence with mutual information
  and redundancy. These bounds are relevant for linking Zipf's law
  with the neural scaling law for large language models. Third, we
  develop a framework for universal codes with fixed infinite
  vocabularies, recasting universal coding as matching ranked patterns
  that are independent of empirical data. Finally, we analyze
  grammar-based codes with finite vocabularies being empirical rank
  lists, proving that that such codes are also universal.  These
  results extend foundations of universal grammar-based coding and
  reaffirm previously stated connections to power laws for human
  language and language models.
%
  \\[1ex]
\end{abstract}


\end{titlepage}
\pagestyle{plain}   


\section{Introduction}
\label{secIntroduction}

Grammar-based codes, a certain natural and successful approach to
universal coding, are a two-step method
\cite{KiefferYang00,KiefferYang22}. First, the input string is
represented as a straight-line grammar, i.e., a context-free grammar
that generates the input string as its sole production. Second, the
straight-line grammar is encoded as the output binary string using a sort
of arithmetic coding. The mapping from the input string to the
straight-line grammar is called the grammar transform, whereas the
mapping from the straight-line grammar to the output binary string is
called the grammar encoder.

The essential idea of grammar-based coding comes from experiments in
computational linguistics that can be traced at least back to
1980's--1990's \cite{Wolff80, DeMarcken96, NevillManning96,
  NevillManningWitten97, NevillManningWitten97b}. Following later
development, grammar transforms are applied for text tokenization in
modern large language models as the byte-pair encoding (BPE)
\cite{SenrichHaddowBirch16, BrownOthers20}.  By contrast, in the
context of information theory, basic understanding of grammar-based
coding bears to the work by Kieffer and Yang \cite{KiefferYang00}. In
particular, Kieffer and Yang demonstrated that a large class of codes
with irreducible grammar transforms and a fixed particular grammar
encoder is universal. Another seminal work by Charikar et al.\
\cite{CharikarOthers05} showed the NP-hardness of computing the
minimal grammar transform, whose length was defined as the sum of
grammar rule lengths.

Our modest contribution to this field \cite{Debowski07c, Debowski11b}
was to consider a local grammar encoder, which encodes grammars symbol
by symbol---in a way that seems naive and suboptimal. We also
considered minimal grammar-based codes with respect to the local
grammar encoder, i.e., grammar transforms whose length after binary
encoding is minimal, given some additional constraints on the classes
of grammars.  Obviously, the local grammar encoder is far from being
optimal, whereas minimal grammars are likely intractable. Why should
we consider them then?

The main motivation for considering such minimal codes is a simple
algebraic upper bound for their pointwise mutual information in terms
of the vocabulary size, i.e., the number of rules in the minimal
grammar \cite{Debowski06, Debowski11b}.  It follows hence that if the
minimal code is universal and the Shannon mutual information or the
algorithmic mutual information for a given process are large then the
minimal grammar must have numerous rules.  Consequently, we may think
that the minimal code may be used for detecting and quantifying the
amount of word-like tokens in the stream of symbols generated by an
information source \cite{Debowski11b}, see also \cite{Wolff80,
  DeMarcken96, NevillManning96, NevillManningWitten97,
  NevillManningWitten97b}. In particular, such a code might be used to
explain or connect certain empirical power laws in linguistics such as
Zipf's law, Heaps' law, and Hilberg's law \cite{Debowski06,
  Debowski11b}.

To be more verbose, Zipf's law states that the rank of a word is
inversely proportional to its frequency \cite{Estoup16, Condon28,
  Zipf35, Mandelbrot54, Baayen01} and Heaps' law states that the
number of distinct words in a text grows like a power of the text
length \cite{Herdan64, Heaps78}.  By contrast, Hilberg's law
conjectures that the mutual information between two adjacent texts of
the same length is also proportional to a power of the text length
\cite{Hilberg90, CrutchfieldFeldman03, TakahiraOthers16,
  HahnFutrell19, TanakaIshii21}. More recently, Hilberg's law has been
known as the neural scaling law for large language models
\cite{HestnessOther17, KaplanOther20,
  HenighanOther2020, HernandezOther21}. Whereas the linguistic reality
is somewhat more complicated than simple power laws
\cite{FerrerSole01b, MontemurroZanette02, Fan10, FontClosCorral15,
  Debowski25}, in the first approximation, we may suppose that minimal
local grammar-based coding combined with Hilberg's law explains Heaps'
law \cite{Debowski11b}.  There are also converse theoretical
results. Given some form of Zipf's law, we obtain Hilberg's law as
well, see \cite{Debowski09, Debowski11b, Debowski21, Debowski21b,
  Hutter21, Debowski23} for the accompanying mathematical framework.

The aim of this paper is to critically update articles
\cite{Debowski07c, Debowski11b} after over one decade. As our
understanding has progressed, we provide an assortment of results that
strengthen or simplify some findings of \cite{Debowski07c,
  Debowski11b}.  Our results are of a theoretical nature. They can be
split into four paths.
\begin{itemize}

\item First of all, in Section \ref{secMinimalCodes}, we notice an
  extremely simple harmonic bound for ranked probabilities, namely,
  $\pi_n\le n^{-1}$, where $\pi_n$ is the $n$-th largest probability
  \mbox{\cite[Eqs. (12) and (43)]{Debowski25}}. This bound reminds of
  Zipf's law $\pi_n\propto n^{-1}$ discussed in quantitative
  linguistics \cite{Estoup16, Condon28, Zipf35}.  Our contribution is
  to apply the harmonic bound to information theory, precisely, to
  universal source coding.  In particular, we strengthen the proof of
  universality of the minimal code from \cite[Theorem 1]{Debowski07c}
  and \cite[Theorem 5]{Debowski11b}. The original proof rested on
  universality of the minimal block code discussed by Neuhoff and
  Shields \cite{NeuhoffShields98}. It should be noted that the
  Neuhoff-Shields code can be interpreted as a grammar-based code but
  it does not apply a local grammar encoder. Moreover, it should be
  noted that the proof in \cite{Debowski07c, Debowski11b} was not
  fully general since it applied only to processes with a strictly
  positive entropy rate. Here we show that the minimal code is
  universal in any case. We supply a simple proof that applies both
  block coding and the harmonic bound, to be clear.

\item As for the second kind of contributions, in Section
  \ref{secVocabularyBounds}, we improve bounds for minimal grammars in
  terms of their vocabulary size stated in \cite[Theorem
  2]{Debowski07c} and \cite[Theorem 6]{Debowski11b}. Besides the known
  upper bound for the mutual information, we produce a new lower bound
  for the redundancy in terms of an intersection of
  vocabularies. Together, these bounds establish sandwich bounds for
  power-law rates of three quantities: the vocabulary size of the
  minimal grammar, the code-based mutual information, and the
  redundancy of the code.  Moreover, we show that the vocabulary size
  of the minimal grammar is lower bounded by a false power law that is
  proportional to the entropy rate of the source. This result extends
  our earlier results \cite[Theorem 4]{Debowski07c} for a somewhat
  different notion of a minimal grammar.  Thus, probing empirically
  the power-law growth of mutual information by means of the minimal
  grammar-based code can be misleading: The mutual information of the
  source may dominate the false power law only for sufficiently long
  data and sufficiently strongly dependent sources.

\item The third kind of results, in Section
  \ref{secInfiniteVocabularies}, is to develop the theory of codes
  with a fixed infinite vocabulary. We show that such codes, called
  wrapped codes, are universal if the vocabulary is obtained by
  sorting strings according to a certain universal code length.  In
  other words, such vocabularies are rank lists with respect to the
  corresponding incomplete distribution.  Here the harmonic bound
  $\pi_n\le n^{-1}$ is also used. These results are mostly of a
  theoretical importance. They challenge the view that universal
  coding is a problem of acquiring some knowledge from the empirical
  data. We show that universal coding can be equivalently regarded as
  a problem of matching pre-learned mathematical patterns in the
  empirical data.  This is not so deep as it may sound but we think
  that this view of universal coding may inspire a more practical
  construction in the future.

\item The fourth bundle of results, in Section
  \ref{secFiniteVocabularies}, scales down the concept of universal
  codes with fixed infinite vocabularies. We want to obtain finite
  empirically driven rank lists that also define universal codes. For
  this goal, we investigate what happens when we restrict an infinite
  universal vocabulary to a finite prefix, obtaining so called trimmed
  codes. Furthermore, we apply these trimmed codes to the empirical
  rank list. We show that a trimmed code based on sorted substrings is
  also universal. This code resembles somewhat the complicated
  algorithm by de Marcken \cite{DeMarcken96}, which was one of the
  original impulses that motivated the field of grammar-based
  coding. Nowadays similar algorithms are used for text tokenization
  in large language models, viz.\ the byte-pair encoding (BPE)
  \cite{SenrichHaddowBirch16}. We suspected that de Marcken's involved
  algorithm is universal and now we can show that a similar code is,
  indeed.
  
\end{itemize}
The article is concluded in Section \ref{secConclusion}. In Appendix
\ref{secCriteria}, we recall the proofs of two effective criteria of
code universality.


\section{Minimal codes}
\label{secMinimalCodes}

In this section, we recall the construction of local grammar-based
codes and we prove their universality. We strengthen, generalize, and
simplify some results exposed in \cite{Debowski07c, Debowski11b}.
Section \ref{secAdmissibleCode} introduces minimal local grammar-based
codes.  In Section \ref{secUniversality}, we demonstrate the
universality of such codes.

\subsection{Straight-line grammars}
\label{secAdmissibleCode}

We denote the set of natural numbers $\mathbb{N}:=\klam{1,2,3,...}$
and integers $\mathbb{Z}:=\klam{...,-1,0,1,...}$.  We also denote the
ordinal infinity $\omega$ and we consider the standard total order $<$
on set $\mathbb{Z}\cup\klam{\omega}$.  In the following, we adopt a
somewhat unusual but algebraically motivated convention. Natural
numbers will be reserved for secondary non-terminal symbols, $\omega$
will stand for the start symbol, $0$ will denote a comma character,
whereas the set of terminal symbols (the input alphabet) will be
formally identified with an interval
$\mathbb{X}:=\klam{-m,-m+1,...,-1}$.  Notation $\mathbb{Y}^n$ denotes
the set of strings over an alphabet $\mathbb{Y}$ of length $n$, the
length itself is defined as $\abs{w}:=n$ for
$w\in\mathbb{Y}^n$. Symbol $\lambda$ denotes the empty string, whereas
$\mathbb{Y}^0:=\klam{\lambda}$. Set
$\mathbb{Y}^{<n}:=\bigcup_{k=0}^{n-1}\mathbb{Y}^k$ is the set of
strings strictly shorter than $n$, and
$\mathbb{Y}^*:=\bigcup_{k=0}^\infty\mathbb{Y}^k$ is the Kleene star.
Strings consisting of individual symbols are abbreviated as
$y_j^k:=(y_j,y_{j+1},...,y_k)$, where $j\le k$ and $y_i\in\mathbb{Y}$.
Notation $f|_A: A\in x\mapsto f(x)$ denotes the restriction of a
function $f$ to a subdomain $A$.

The definition of a straight-line grammar, called an admissible
grammar in \cite{KiefferYang00}, can be stated succinctly as
follows. We assume that the grammar can be infinite, which will matter
for some later considerations.
\begin{definition}[straight-line grammar]
  A vocabulary is a function
  \begin{align}
    S:V(S)\to(\mathbb{X}\cap V(S))^*
  \end{align}
  where $V(S)=\klam{1,2,...,v}$ or $V(S)=\mathbb{N}$ and for every
  $S(r)=(r_1,r_2,...,r_p)$ we have $r_i<r$.  Strings $S(r)$ are called
  secondary rules.  Cardinality $\card V(S)$ is called the
  vocabulary size.  A primary rule for a vocabulary $S$ is a string
  $P\in(\mathbb{X}\cap V(S))^*$. A straight-line grammar with a
  vocabulary $S$ and a primary rule $P$ for $S$ is a function
  \begin{align}
    \label{Grammar}
    G:V(S)\cup\klam{\omega}\to(\mathbb{X}\cap V(S))^*
  \end{align}
  where $G|_{V(S)}=S$ and $G(\omega)=P$. The class of straight-line
  grammars is denoted $\mathcal{G}$.
\end{definition}
In the above, as we said, natural numbers correspond to non-terminal
symbols, whereas the negative numbers correspond to terminal symbols.
There is exactly one rule $G(r)$ per non-terminal symbol $r$ and each
non-terminal symbol $r$ can be rewritten only onto smaller symbols.

The production of a string by a straight-line grammar can be also made
precise in a simple way in the next definition.
\begin{definition}[grammar expansion]
  For a vocabulary $S$, we
  iteratively define its expansion function
  \begin{align}
    \label{Expansion}
    S^*:\mathbb{X}\cup V(S)\to\mathbb{X}^*
  \end{align}
  as $S^*(r):=r$ for $r\in\mathbb{X}$ and concatenation
  $S^*(r):=S^*(r_1)S^*(r_2)...S^*(r_p)$ for $S(r)=(r_1,r_2,...,r_p)$.
  The expansion of a primary rule $P$ for a vocabulary $S$ is
  $P^*:=S^*(r_1)S^*(r_2)...S^*(r_p)$ for $P=(r_1,r_2,...,r_p)$.  We
  say that a straight-line grammar $G$ produces a string
  $u\in\mathbb{X}^*$ if $P^*=u$ holds for its primary rule.
\end{definition}

There are a few interesting subclasses of straight-line grammars.
\begin{definition}[finite grammar]
  A straight-line grammar $G$ with a vocabulary $S$ is called finite
  if $V(S)$ is finite and infinite if $V(S)$ is infinite.
\end{definition}
\begin{definition}[flat grammar]
  A vocabulary $S$ is called flat if $S:V(S)\to\mathbb{X}^*$. A
  straight-line grammar is called a flat grammar if its vocabulary is
  flat. The class of flat grammars is denoted $\mathcal{F}$.
\end{definition}
\begin{definition}[block grammar]
  \label{defiBlockGrammar}
  A vocabulary $S$ is called $k$-bloc if $S:V(S)\to\mathbb{X}^k$. A
  straight-line grammar $G$ is called a $k$-block grammar if its
  vocabulary $S$ is $k$-block and its primary rule $P$ satisfies
  $P\in\mathbb{X}^{<k}\times V(S)^*\times \mathbb{X}^{<k}$.  A
  straight-line grammar is called a block grammar if it is a $k$-block
  grammar for a certain $k$. The class of block grammars is denoted
  $\mathcal{B}$.
\end{definition}
\emph{Remark:} In \cite{Debowski07c, Debowski11b}, a $k$-block grammar
$G$ was defined as a straight-line grammar with a $k$-block vocabulary
$S$ and a primary rule $P\in V(S)^*\times \mathbb{X}^{<k}$. Using
$P\in\mathbb{X}^{<k}\times V_G^*\times \mathbb{X}^{<k}$ rather than
$P\in V_G^*\times \mathbb{X}^{<k}$ is crucial to apply the
universality criterion (\ref{UniversalityCriterionBlock}) discussed in
the next section. This petty difference allows to circumvent annoying
theoretical problems with the distinction between ergodicity and
$k$-ergodicity \cite{Gray09}.
\begin{definition}
  A subclass $\mathcal{J}$ of straight-line grammars is called
  sufficient if for each string there exists a grammar in
  $\mathcal{J}$ which produces this string
\end{definition}
In particular, classes $\mathcal{G}$, $\mathcal{F}$, and $\mathcal{B}$
are sufficient.
\begin{definition}[minimal grammar transform]
  Let $\mathcal{J}$ be a sufficient subclass of straight-line
  grammars. Let $\norm{\cdot}:\mathcal{J}\to[0,\infty]$ be a real
  function.  We define the minimal grammar transform
  $\Gamma^{\mathcal{J}}_{\norm{\cdot}}(u)$ as the grammar
  $G\in\mathcal{J}$ that produces string $u\in\mathbb{X}^*$ and
  minimizes function $\norm{G}$.
\end{definition}

There is a simple grammar length function
\begin{align}
  \label{YangKieffer}
  \abs{G}:=\sum_{r\in V(S)\cup\klam{\omega}}\abs{G(r)}.
\end{align}
This length is considered as the default length of a grammar in the
approaches of \cite{KiefferYang00} and \cite{CharikarOthers05}. We
call (\ref{YangKieffer}) the Yang-Kieffer length of grammar $G$.  It
was demonstrated in \cite{CharikarOthers05} that computing the
unconstrained minimal grammar transform with respect to the
Yang-Kieffer length $\Gamma^{\mathcal{G}}_{\abs{\cdot}}$ is
NP-hard. We suppose that computing unconstrained minimal transforms
for other non-trivial length functions is equally hard.

In papers \cite{Debowski07c, Debowski11b}, we introduced a local
grammar encoder, which encodes straight-line grammars symbol by symbol,
with additional two symbols for commas.  The encoder applied relative
indexing of nonterminal symbols and the encoder was prefix-free. Here
we give up the idea of relative indexing. Leading to simpler
notations, this will have no effect on universality and a few other
properties of grammar-based codes. As a result, we define the
local grammar encoder as follows. 
\begin{definition}[local grammar encoder]
  Consider a prefix-free code for restricted integers
  $\psi:\mathbb{X}\cup\klam{0}\cup\mathbb{N}\to\klam{0,1}^*$. The
  local grammar encoder $\psi^*$ for a finite straight-line grammar
  $G$ with a primary rule $P$ and a vocabulary $S$ returns string
  \begin{align}
    \psi^{**}(G):=\psi(\card V(S))
    \psi^*(S(1))\psi^*(S(2))...\psi^*(S(\card V(S)))\psi^*(P),
  \end{align}
  where
  $\psi^*((r_1,r_2,...,r_p)):=\psi(r_1)\psi(r_2)...\psi(r_p)\psi(0)$.
\end{definition}

Consequently, in \cite{Debowski07c, Debowski11b}, we investigated the
minimal grammar transform that minimizes the grammar length defined
via the local grammar encoder.
\begin{definition}[minimal code]
  Let $\mathcal{J}$ be a sufficient subclass of straight-line
  grammars. Let $\psi^*$ be a local grammar encoder. We define the
  $\psi$-minimal grammar transform
  $\Gamma^{\mathcal{J}}_\psi(u):=\Gamma^{\mathcal{J}}_{\norm{\cdot}}(u)$
  with respect to length $\norm{G}=\abs{\psi^{**}(G)}$. Subsequently, the
  $\psi$-minimal code
  $B^{\mathcal{J}}_\psi:\mathbb{X}^*\to\klam{0,1}^*$ is defined as
  \begin{align}
    \label{MinimalAdmissibleCode}
    B^{\mathcal{J}}_\psi(u)=\psi^*(\Gamma^{\mathcal{J}}_\psi(u)).
  \end{align}
\end{definition}
As we remarked, finding the minimal straight-line grammar for a given
string may be intractable because it requires searching globally
through a prohibitively large space of straight-line grammars.

The minimal codes depend on the choice of the local encoder $\psi^*$.
Let $\log r$ denote the binary logarithm of $r$. We will consider the
following definition.
\begin{definition}[proper code]
  A code $\psi:\mathbb{X}\cup\klam{0}\cup\mathbb{N}\to\klam{0,1}^*$ is
  called proper if
  \begin{enumerate}
  \item $\psi$ is prefix-free;
  \item $\abs{\psi(r+1)}\ge\abs{\psi(r)}$ for $r\in\mathbb{X}\cup
    \klam{0}\cup\mathbb{N}$;
  \item $\abs{\psi(r)}=c_1$ for $r\in\mathbb{X}\cup\klam{0}$;
  \item $\abs{\psi(r)}\le\log r+2\log(\log r+1)+c_2$ for
    $r\in\mathbb{N}$ and some $c_2<\infty$.
  \end{enumerate}
  Succinctly, $\psi$-minimal grammars and codes with a proper
  code $\psi$ are called proper.
\end{definition}
Conditions 2 and 3 secure a simpler behavior of the minimal
grammar. For example, condition 2 causes that the order of optimal
secondary rules in the grammar is biased towards the empirical ranking
of strings. This eases computing a constrained minimum of the grammar
length. By contrast, condition 3 leads to simpler bounds for the
length of the grammar-based code and the code-based mutual
information. 

Proper codes exist by the Kraft inequality. In particular, there
exists a proper code $\psi$ of form 
\begin{align}
  \label{ExactPsi}
  \psi(r):=
  \begin{cases}
    \phi(r)
    ,
    &
    r\in\mathbb{X}\cup\klam{0}
    ,
    \\
    \phi(1)\delta(r),
    &
    r\in\mathbb{N}
    ,
  \end{cases}
\end{align}
where $\phi:\mathbb{X}\cup\klam{0,1}\to\klam{0,1}^*$ is a fixed-length code
with length
\begin{align}
  \abs{\phi(n)}=1+\floor{\log(m+2)}
\end{align}
and $\delta:\klam{1,2,...}\to\klam{0,1}^*$ is the Elias delta code
\cite{Elias75} with length
\begin{align}
  \abs{\delta(n)}=\floor{\log n}+2\floor{\log(\floor{\log n}+1)}+2.
\end{align}

In \cite{Debowski07c, Debowski11b}, we proved that proper codes
$B^{\mathcal{G}}_\psi$, $B^{\mathcal{F}}_\psi$, and
$B^{\mathcal{B}}_\psi$ are universal for stationary ergodic processes
with a strictly positive entropy rate. Block grammars were also
considered by Neuhoff and Shields \cite{NeuhoffShields98}. They used
them to construct a certain universal minimal block code which does
not apply the local grammar encoder.

In contrast to codes $B^{\mathcal{G}}_\psi$ and
$B^{\mathcal{F}}_\psi$, which may be NP-hard to compute, the
proper minimal block code $B^{\mathcal{B}}_\psi$ can be provably
computed in a time close to linear (with some logarithmic
add-ons). For this goal, we have to consider all parsings of the input
string into $k$-blocks and to minimize the code length over $k$. To
determine the optimal code length for each of these parsings, we
notice that by inequality $\abs{\psi(r+1)}\ge\abs{\psi(r)}$, the
optimal secondary rules should be sorted according to the ranked
empirical distribution of $k$-blocks.  Once such sorting is performed,
the resulted code is minimal within the class of block grammars since
all rules have the same length after local encoding by equality
$\abs{\psi(r)}=c_1$ for $r\in\mathbb{X}\cup\klam{0}$.

Actually, when writing papers \cite{Debowski07c, Debowski11b}, we did
not notice that the proper minimal block code can be so easily
computed. For this reason, we did not include any numerical experiment
in \cite{Debowski07c, Debowski11b}.

\subsection{Universality}
\label{secUniversality}

Since the proper minimal block code is uniquely decodable and it
achieves a local minimum, we can demonstrate easily that this code is
universal. We briefly recall what it means.  Let us denote the entropy
rate of a stationary process $(X_i)_{i\in\mathbb{Z}}$ over alphabet
$\mathbb{X}$ as
\begin{align}
  h:=\lim_{n\to\infty} \frac{H(X_1^n)}{n},
\end{align}
where $H(X):=\mean\kwad{-\log P(X)}$ is the Shannon entropy of random
variable $X$. A function $Q:\mathbb{X}^*\to[0,1]$ is called an
incomplete distribution if we have the Kraft inequality
$\sum_{u\in\mathbb{X}^*} Q(u)\le 1$.
\begin{definition}[adapted and universal distributions]
  An incomplete distribution $Q$ is called adapted to a stationary
  ergodic process $(X_i)_{i\in\mathbb{Z}}$ over alphabet
  $\mathbb{X}$ if
  \begin{align}
    \lim_{n\to\infty} \frac{\kwad{-\log Q(X_1^n)}}{n}
    &= h \text{
    a.s.},
    \\
    \lim_{n\to\infty} \frac{\mean\kwad{-\log Q(X_1^n)}}{n}
    &= h.
  \end{align}
  An incomplete distribution $Q$ is called universal if it is adapted
  to any stationary ergodic process $(X_i)_{i\in\mathbb{Z}}$ over
  alphabet $\mathbb{X}$.
\end{definition}

We note that universal distribution exist only for finite alphabets
\cite{GyorfiPaliMeulen94}.  There are two important cases of
incomplete distributions: First, let $B:\mathbb{X}\to\klam{0,1}^*$ be
a prefix-free code. By the Kraft inequality, function
$Q_B(u):=2^{-\abs{B(u)}}$ is an incomplete distribution.  We call code
$B$ universal if $Q_B$ is universal. Second, let $K(u)$ be the
prefix-free Kolmogorov complexity of string $u$ \cite{Solomonoff64,
  Kolmogorov65en2, Chaitin75, LiVitanyi08}.  Function
$Q_K(u):=2^{-K(u)}$ is also an incomplete universal distribution.

There is a simple universality criterion which is based on the $k$-th
order conditional empirical entropy. We call it the conditional
criterion to distinguish it from another criterion which we will
introduce subsequently.
\begin{theorem}[conditional universality criterion,
  \cite{ZivLempel77,KiefferYang00}]
  \label{theoCriterionMarkov}
  An incomplete distribution $Q$ is universal if for any $k\ge 1$, any
  conditional probability distribution
  $\pi:\mathbb{X}\times\mathbb{X}^k\to[0,1]$, and any
  string $x_1^n\in\mathbb{X}^*$, we have
  \begin{align}
    \label{UniversalityCriterionMarkov}
    -\log Q(x_1^n)\le C(n,k)-\log\prod_{i=k+1}^{n}\pi(x_i|x_{i-k}^{i-1}),
  \end{align}
  where $\lim_{k\to\infty} \limsup_{n\to\infty} C(n,k)/n=0$.
\end{theorem}

The conditional universality criterion was first applied in the
celebrated Ziv inequality for proving universality of the Lempel-Ziv
code \cite{ZivLempel77,CoverThomas06}.  Moreover, in
\cite{KiefferYang00, OchoaNavarro19} it was shown that the $k$-th
order conditional empirical entropy bounds the lengths of
grammar-based codes that apply the encoder by Kieffer and Yang
\cite{KiefferYang00}. Such an inequality is sufficient to assert
universality of the respective codes.

In contrast, for our purpose, we need a somewhat different
universality criterion, which is based on the $k$-block unconditional
empirical entropy. Thus we call this criterion the block criterion.
\begin{theorem}[block universality criterion]
  \label{theoCriterionBlock}
  An incomplete distribution $Q$ is universal if for any
  $k\ge 1$, any probability distribution $\pi:\mathbb{X}^k\to[0,1]$,
  and any string $x_1^n\in\mathbb{X}^*$, we have
  \begin{align}
    \label{UniversalityCriterionBlock}
    -\log Q(x_1^n)\le C(n,k)-\frac{1}{k}\log
    \prod_{i=0}^{n-k}\pi(x_{i+1}^{i+k}),
  \end{align}
  where $\lim_{k\to\infty} \limsup_{n\to\infty} C(n,k)/n=0$.
\end{theorem}
Theorems \ref{theoCriterionMarkov} and \ref{theoCriterionBlock} belong
to the toolbox or folklore of information theory, respectively. Since
we had troubles in locating the first exact statements of these
criteria in the literature, especially of the block universality
criterion, we present their short proofs for a meaningful comparison
in Appendix \ref{secCriteria}.

An interesting detail is that the right-hand side of criterion
(\ref{UniversalityCriterionBlock}) contains probabilities of
overlapping blocks. If there were no overlaps and no division by $k$,
we might encounter problems with the distinction between ergodicity
and $k$-ergodicity \cite{Gray09}. Fortunately, for the successful
application of criterion (\ref{UniversalityCriterionBlock}) to block
codes, it is sufficient to consider all $k$ distinct shifts of blocks
and take the shift that yields the shortest code.

Exactly this idea is used in the following Theorem \ref{theoUniversal}
for proving universality of the proper minimal block code.  We may
prove that the $\psi$-minimal block code is universal for any proper
code $\psi$ for extended natural numbers.  Our result generalizes and
simplifies the earlier results \cite[Theorem 1]{Debowski07c} and
\cite[Theorem 5]{Debowski11b}. It is as follows.
\begin{theorem}
  \label{theoUniversal}
  The proper minimal block code $B^{\mathcal{B}}_\psi$ is
  universal.
\end{theorem}
The proof is presented after the proof of the auxiliary Lemma
\ref{theoHarmonicBound}. In the previous exposition \cite[Theorem
5]{Debowski11b}, we only proved universality for processes with $h>0$,
whereas some earlier result \cite[Theorem 1]{Debowski07c} was stated
mistakenly without this assumption.

Obviously codes $B^{\mathcal{G}}_\psi$ and $B^{\mathcal{F}}_\psi$ are
shorter than the minimal block code $B^{\mathcal{B}}_\psi$. In
consequence, universality of these codes follows by the universality
of the proper minimal block code. By means of our new proof of
Theorem \ref{theoUniversal}, it will also become obviously clear why
some variations on the theme of the minimal block code discussed by
Neuhoff and Shields \cite{NeuhoffShields98} are also universal.  The
key observation is the following extremely simple inequality which
implies that ranked probabilities are upper bounded by the harmonic
series.
\begin{lemma}[harmonic bound, cf. \mbox{\cite[Eqs. (12) and
  (43)]{Debowski25}}]
  \label{theoHarmonicBound}
  Let $\pi_1\ge\pi_2\ge ...$ be a sequence of probabilities of
  disjoint events, i.e., $\sum_{i} \pi_i\le 1$. Then
  \begin{align}
    \label{HarmonicBound}
    \pi_n\le \frac{1}{n}.
  \end{align}
\end{lemma}
\begin{proof}
  We have $n\pi_n\le \sum_{j=1}^n \pi_j\le 1$. Dividing by $n$, we
  obtain (\ref{HarmonicBound}).
\end{proof}

Lemma \ref{theoHarmonicBound} reminds of Zipf's law
$\pi_n\propto n^{-1}$ for the rank-frequency distribution of words in
natural language \cite{Estoup16, Condon28, Zipf35}. By Lemma
\ref{theoHarmonicBound}, if we use a local grammar-based code where
rules $G(n)$ are sorted by arbitrary probabilities $\pi_n$ then the
binary identifier $\psi(n)$ of the $n$-th non-terminal will be roughly
shorter than the respective minus log-probability $-\log \pi_n$ if we
have $\abs{\psi(n)}\approx \log n$ in general.  This prompts a path to
proving universality of the $\psi$-minimal block code. We note that
the calculations below and in the paper by Neuhoff and Shields
\cite{NeuhoffShields98} apply different ideas and take distinct paths,
our reasoning being simpler.
\begin{proof*}{Theorem \ref{theoUniversal}}
  It suffices to show that the $\psi$-minimal block code satisfies
  universality criterion (\ref{UniversalityCriterionBlock}).  We will
  consider a sequence of $k$-block grammars $G_l$ for string $x_1^n$
  indexed by index $l\in\klam{0,1,...,k-1}$ such that:
  \begin{itemize}    
  \item The secondary rules, regardless of index $l$, define all
    $k$-blocks in the order of ranking given by the distribution
    $\pi$:
    \begin{align}
      G_l(r)\in\mathbb{X}^k
      \text{ for } r\in\klam{1,2,...,m^k} \text{ and }
      \pi(G(r))\ge\pi(G(r+1))
      .
    \end{align}
  \item The primary rule of each grammar $G_l$ defines string $x_1^n$
    using the identifiers for $k$-blocks shifted by $l$ positions:
    \begin{align}
      G_l(\omega)=(R_1,R_2,...,R_l,r_1^l,r_2^l,...,r_{p_l}^l,
      R_{-l'_l},R_{-l'_l+1},...,R_{-1})
    \end{align}
    where $R_i\in\mathbb{X}$, $r_i^l\in\klam{1,2,...,m^k}$, and
    $0\le l,l'_l<k$.
  \end{itemize}
  We observe that none of these grammars can be better than the
  $\psi$-minimal block grammar for $x_1^n$. Hence, for any
  $l\in\klam{0,1,...,k-1}$, we may bound
  \begin{align}
    \abs{B^{\mathcal{B}}_\psi(x_1^n)}
    &\le
      \abs{\psi^*(G_l)}
      \le
      C(k)
      +\sum_{i=1}^{p_l}\abs{\psi(r_i^l)}
      ,
  \end{align}
  where
  $C(k):=\kwad{m^k(k+1)+2k+2}\abs{\psi(0)}$.

  We have inequality $\abs{\psi(r)}\le \log r+2\log(\log r+1)+c_2$
  for $r\in\mathbb{N}$ by the hypothesis and inequality
  $\pi(G(r))\le 1/r$ by Lemma \ref{theoHarmonicBound}. Hence,
  we may further bound
  \begin{align}
    \sum_{i=1}^{p_l}\abs{\psi(r_i^l)}
    &\le
      \sum_{i=1}^{p_l}\kwad{\log r_i^l +2\log(\log m^k+1)+c_2}
      \nonumber\\
    &\le
      \frac{n}{k}
      \kwad{2\log(\log m^k+1)+c_2}
      -\sum_{i=1}^{p_l}\log\pi(G(r_i^l))
      .
  \end{align}
  Denote $C(n,k):=C(k)+\frac{n}{k}\kwad{2\log(\log m^k+1)+c_2}$. Then
  we may bound
  \begin{align}
    \abs{B^{\mathcal{B}}_\psi(x_1^n)}
    &\le
      C(n,k)
      -
      \min_{l\in\klam{0,1,...,k-1}}
      \sum_{i=1}^{p_l}\log\pi(G(r_i^l))
    \nonumber\\
    &\le
      C(n,k)
      -
      \frac{1}{k}\sum_{l=0}^{k-1}
      \sum_{i=1}^{p_l}\log\pi(G(r_i^l))
    \nonumber\\
    &=
      C(n,k)
      -
      \frac{1}{k}
      \sum_{i=0}^{n-k}\log\pi(x_{i+1}^{i+k})
      .
  \end{align}
  To conclude, we observe $\lim_{k\to\infty} \limsup_{n\to\infty} C(n,k)/n=0$.
\end{proof*}

\section{Vocabulary bounds}
\label{secVocabularyBounds}

How can the proper minimal codes be useful if the local grammar
encoding is so suboptimal?  The most important motivation for these
codes is a bound for their pointwise mutual information in terms of
the number of rules in the proper minimal grammar.  In particular, if
the proper minimal codes are universal and the Shannon mutual
information or the algorithmic mutual information for a given process
is large then the proper minimal grammars must have many rules. Thus,
we may be tempted to think that proper minimal grammars---or similar
ones---might be used for quantifying simultaneously the amount of
memory and the amount of structure in the realization of a stochastic
process. In particular, they can be used for explaining Zipf's law for
natural language via Hilberg's law, as mentioned in Section
\ref{secIntroduction}.

In this section we will state the relevant results that strengthen
somewhat the exposition of \cite{Debowski07c, Debowski11b} and
simplify the notation. Namely, in Section \ref{secMutualInformation},
we introduce basic bounds for mutual information.  In Section
\ref{secFalsePowerLaw}, we derive the false power-law growth of the
minimal straight-line vocabulary. Section \ref{secOptimalBlockLength}
presents a bound for the common length of rules in the minimal block
code. We note in passing that our new notation will inspire some more
novel developments of Section \ref{secInfiniteVocabularies}.
We assume a proper code $\psi$.

\subsection{Mutual information}
\label{secMutualInformation}

Let us introduce some auxiliary concepts and notations.  A vocabulary
$S$ is called finite if $\card V(S)<\infty$ and infinite if
$\card V(S)=\infty$.  For a grammar $G$ with a primary rule $P$ and a
finite vocabulary $S$, we write $G=S\oplus P$.  In this case, we also
write
\begin{align}
  \norm{G}&:=\norm{P}+\norm{S}+\norm{\card V(S)},
  \\
  \norm{P}&:=\abs{\psi^*(P)},
  \\
  \norm{S}&:=\sum_{r\in V(S)} \abs{\psi^*(S(r))},
  \\
  \norm{r}&:=\abs{\psi(\ceil{r})}.
\end{align}
We define the
diameter of a vocabulary $S$ as
\begin{align}
  d(S):=
  \sup_{r\in V(S)} \abs{\psi^*(S^*(r))}.
\end{align}
Obviously, $d(S)\le\norm{S}$ if $S$ is flat (or block, in particular).
We also write
\begin{align}
  \norm{S}_+&:=\norm{S}+\norm{\card V(S)},
  \\
  \norm{S}_-&:=\norm{S}-\norm{\card V(S)}.
\end{align}
We have $\card V(S)\le\norm{S}$.

For the minimal grammar $\Gamma^{\mathcal{J}}_\psi(u)$, we denote its
vocabulary as $\Sigma^{\mathcal{J}}_\psi(u)$ and its vocabulary size
as $\card V(\Sigma^{\mathcal{J}}_\psi(u))$. We also define the minimal
primary rule $\Pi^{\mathcal{J}}_\psi(u|S)$ as a string
$P\in\mathbb{Z}^*$ such that grammar $G$ with primary rule $P$ and
vocabulary $S$ belongs to $\mathcal{J}$ and minimizes length
$\norm{G}$. Obviously, we have
\begin{align}
  \label{DecompositionAdmissible}
  \norm{\Gamma^{\mathcal{J}}_\psi(u)}
  &=\norm{\Pi^{\mathcal{J}}_\psi(u|\Sigma^{\mathcal{J}}_\psi(u))}
    +\norm{\Sigma^{\mathcal{J}}_\psi(u)}_+.
\end{align}
The minimal primary rules $\Pi^{\mathcal{J}}_\psi(u|S)$ for
$\mathcal{J}\in\klam{\mathcal{G},\mathcal{F},\mathcal{B}}$ and a
finite vocabulary $S$ can be computed in time
$O(\abs{u}\cdot \card V(S))$ via a dynamic programming algorithm that
resembles the Viterbi search for hidden Markov models
\cite{Jelinek97}.

Having these notations, we observe an important fact which is
abstracted from the more specific results \cite[Theorem
2]{Debowski07c} and \cite[Theorem 6]{Debowski11b}.
\begin{theorem}
  \label{theoMIPrimary}
  For $\mathcal{J}\in\klam{\mathcal{G},\mathcal{F},\mathcal{B}}$, we
  have
  \begin{align}
    \label{MIPrimaryUpper}
    \norm{\Pi^{\mathcal{J}}_\psi(u|S)}
    +
    \norm{\Pi^{\mathcal{J}}_\psi(v|S)}
    -
    \norm{\Pi^{\mathcal{J}}_\psi(uv|S)}
    \le
      \norm{S}.
  \end{align}
  For $\mathcal{J}\in\klam{\mathcal{G},\mathcal{F}}$, we
  have
  \begin{align}
    \label{MIPrimaryLower}
    c_1
    \le
    \norm{\Pi^{\mathcal{J}}_\psi(u|S)}
    +
    \norm{\Pi^{\mathcal{J}}_\psi(v|S)}
    -
    \norm{\Pi^{\mathcal{J}}_\psi(uv|S)}.
  \end{align}
\end{theorem}
\begin{proof}
  To prove the upper bound (\ref{MIPrimaryUpper}), we may write
  $\Pi^{\mathcal{J}}_\psi(uv|S)=x_1Z_1y_1, S(Z_1)=x_2Z_2y_2, ...,
  S(Z_{k-1})=x_{k-1}Z_{k-1}y_{k-1}, S(Z_k)=x_ky_k$, whereas
  $S\oplus x_1x_2...x_k$ is a straight-line grammar that produces
  string $u$, and $S\oplus y_ky_{k-1}...y_1$ is a straight-line
  grammar that produces string $v$.  Hence
  \begin{align}
    \norm{\Pi^{\mathcal{J}}_\psi(uv|S)}+\norm{S}
    &\ge \norm{x_1x_2...x_k}+\norm{y_ky_{k-1}...y_1}
    \nonumber\\
    &\ge
    \norm{\Pi^{\mathcal{J}}_\psi(u|S)}+
    \norm{\Pi^{\mathcal{J}}_\psi(v|S)}.
  \end{align}
  Regrouping yields the claim.
  
  To prove the lower bound (\ref{MIPrimaryLower}), we put
  $w:=\Pi^{\mathcal{J}}_\psi(u|S)\Pi^{\mathcal{J}}_\psi(v|S)$. Obviously
  $S\oplus w$ is a straight-line grammar that produces string
  $uv$. Hence
  \begin{align}
    \norm{\Pi^{\mathcal{J}}_\psi(u|S)} + \norm{\Pi^{\mathcal{J}}_\psi(v|S)}
    -c_1
    &=
      \norm{w}
      \nonumber\\
    &\ge
      \norm{\Pi^{\mathcal{J}}_\psi(uv|S)}.
  \end{align}
  Regrouping yields the claim.  The lower bound (\ref{MIPrimaryLower})
  for block grammars ($\mathcal{J}=\mathcal{B}$) is not guaranteed for
  the reason that the concatenation of two block primary rules need
  not be a block primary rule.
\end{proof}

Let us denote the pointwise mutual information
\begin{align} 
  J^{\mathcal{J}}_\psi(u;v)
  &:=\norm{\Gamma^{\mathcal{J}}_\psi(u)}+\norm{\Gamma^{\mathcal{J}}_\psi(v)}-\norm{\Gamma^{\mathcal{J}}_\psi(uv)}.
\end{align}
Theorem \ref{theoMIPrimary} has a corollary that bounds this quantity,
stated in a bit different notation in \cite[Theorem 2]{Debowski07c}
and \cite[Theorem 6]{Debowski11b}.
\begin{theorem}
  \label{theoMISecondary}
  For $\mathcal{J}\in\klam{\mathcal{G},\mathcal{F},\mathcal{B}}$, we
  have
  \begin{align}
    \label{MISecondary}
    J^{\mathcal{J}}_\psi(u;v)
    &\le
      2\norm{\Sigma^{\mathcal{J}}_\psi(uv)}_+
      .
  \end{align}
\end{theorem}
\begin{proof}
  Let $S=\Sigma^{\mathcal{J}}_\psi(uv)$. We have
  \begin{align}
    \norm{\Gamma^{\mathcal{J}}_\psi(u)}
    &\le
      \norm{\Pi^{\mathcal{J}}_\psi(u|S)}+\norm{S}_+,
      \\
    \norm{\Gamma^{\mathcal{J}}_\psi(v)}
    &\le
      \norm{\Pi^{\mathcal{J}}_\psi(v|S)}+\norm{S}_+,
      \\
    \norm{\Gamma^{\mathcal{J}}_\psi(uv)}
    &=
      \norm{\Pi^{\mathcal{J}}_\psi(uv|S)}+\norm{S}_+.
  \end{align}
  Hence (\ref{MISecondary}) follows from
  (\ref{MIPrimaryUpper}). 
\end{proof}

Subsequently, we want to show that the vocabulary size of the minimal
grammar is a proxy for the mutual information.  Let $L(u)$ be the
maximal length of a repetition in $u$
\cite{DeLuca99,KolpakovKucherov99a,KolpakovKucherov99}, namely,
\begin{align}
  \label{MaxRep}
  L(u):=\max\klam{\abs{v}:
  u=xvz=x'vz' \text{ for some } x\neq x', z\neq z'}.
\end{align}
The maximal length of repetition $L(u)$ grows relatively slow in terms
of length $\abs{u}$ for many empirical strings but is not bounded in
terms of the Shannon entropy rate \cite{Shields92b,Shields97}. For
some strongly mixing processes, it is bounded
$L(u)\approx h_2^{-1}\log\abs{u}$, where $h_2$ is the R\'enyi entropy rate
\cite{Szpankowski93,Szpankowski93a}, but this case does not cover the
natural language, where $L(u)\propto (\log\abs{u})^3$
\cite{Debowski12b,Debowski15f,Debowski21}.

Abstracting some specific auxiliary observations in \cite{Debowski07c}
and \cite{Debowski11b}, we may state the following proposition.
\begin{theorem}
  \label{theoMaxRepSecondary}
  For $\mathcal{J}\in\klam{\mathcal{G},\mathcal{F},\mathcal{B}}$, we
  have
  \begin{align}
    \frac{d(\Sigma^{\mathcal{J}}_\psi(u))}{c_1}
    &\le L(u)+1,
    \\
    \card V(\Sigma^{\mathcal{J}}_\psi(u))
    \le
    \frac{\norm{\Sigma^{\mathcal{J}}_\psi(u)}}{c_1}
    &\le
      \card V(\Sigma^{\mathcal{J}}_\psi(u))[L(u)+1],\abs{u}.
  \end{align}
\end{theorem}
\begin{proof}
  Let $\mathcal{J}\in\klam{\mathcal{G},\mathcal{F}}$.  We have
  $d(\Sigma^{\mathcal{J}}_\psi(u))\le c_1[L(u)+1]$ since each
  secondary rule in the minimal straight-line grammar must be used at
  least twice in the primary rule. Obviously,
  $c_1 \card V(\Sigma^{\mathcal{J}}_\psi(u))\le
  \norm{\Sigma^{\mathcal{J}}_\psi(u)}\le c_1
  \card V(\Sigma^{\mathcal{J}}_\psi(u))d(\Sigma^{\mathcal{J}}_\psi(u))$ since each
  rule is written in the minimal straight-line vocabulary in the most
  succinct form. Moreover, we have
  $\norm{\Sigma^{\mathcal{J}}_\psi(u)}\le
  \norm{\Gamma^{\mathcal{J}}_\psi(u)}\le c_1\abs{u}$ by minimality.

  For a bit different reason, we have
  $d(\Sigma^{\mathcal{B}}_\psi(u))\le c_1[L(u)+1]$ because at least
  one secondary rule in the minimal block grammar must be used at
  least twice in the primary rule.  Furthermore,
  $c_1 \card V(\Sigma^{\mathcal{B}}_\psi(u))\le
  \norm{\Sigma^{\mathcal{B}}_\psi(u)}\le c_1
  \card V(\Sigma^{\mathcal{B}}_\psi(u))d(\Sigma^{\mathcal{B}}_\psi(u))$ since each
  rule has the same length and is equal to its expansion in the
  minimal block grammar. Besides,
  $\norm{\Sigma^{\mathcal{B}}_\psi(u)}\le
  \norm{\Gamma^{\mathcal{B}}_\psi(u)}\le c_1\abs{u}$ holds by the
  constrained minimality.
\end{proof}

So far, we have restated known results from
\cite{Debowski07c,Debowski11b}.  Now we will present a novel lower
bound for the redundancies of minimal codes.  For vocabularies $S$ and
$T$, let us define their intersection and difference via
\begin{align}
  S\triangle T&:=\inf\klam{\card V(S)+1}\cup\klam{r\ge 1: S(r)\neq T(r)},
  \\
  S\cap T&:=S|_{\klam{r\ge 1:r<S\triangle T}},
  \\
  S\setminus T&:=S|_{\klam{r\ge 1:r\ge S\triangle T}}.
\end{align}
We can bound the redundancy in this way, which was not noticed in
\cite{Debowski07c,Debowski11b}.
\begin{theorem}
  \label{theoRedundancySecondary}
  Let $\mathcal{J}\in\klam{\mathcal{G},\mathcal{F},\mathcal{B}}$. For
  a stationary process $(X_i)_{i\in\mathbb{Z}}$ over alphabet
  $\mathbb{X}$, we have
  \begin{align}
    \label{RedundancySecondary}
    \mean\norm{\Gamma^{\mathcal{J}}_\psi(X_1^n)}-nh
    &\ge
      \sup_{\tau}
      \mean\norm{\Sigma^{\mathcal{J}}_\psi(X_1^n)\cap \tau(X_{-\infty}^{0})},
  \end{align}
  where $\tau$ ranges through measurable mappings of pasts into
  vocabularies.
\end{theorem}
\begin{proof}
  Denote $S:=\Sigma^{\mathcal{J}}_\psi(X_1^n)$ and
  $T:=\tau(X_{-\infty}^{0})$.  By the conditional
  source coding inequality, we have
  \begin{align}
    \mean\norm{\Pi^{\mathcal{J}}_\psi(X_1^n|S)}
    &\ge
    H(X_1^n|S)
      \ge
      H(X_1^n|S,X_{-\infty}^{0})
     \nonumber\\ 
    &=H(X_1^n|X_{-\infty}^{0})-I(X_1^n;S|X_{-\infty}^{0})
      \nonumber\\
    &\ge
      nh-H(S|X_{-\infty}^{0})
      \ge nh-H(S|T).
  \end{align}
  Also by source coding, we may bound
  \begin{align}
    H(S|T)
    &=H(S\triangle T,S\setminus T|T)
      \nonumber\\
    &\le
      H(S\triangle T,S\setminus T)
      \le
      \mean\kwad{\norm{S}_+-\norm{S\cap T}_-}.
  \end{align}
  Hence we obtain
  \begin{align}
    \mean\norm{\Gamma^{\mathcal{J}}_\psi(X_1^n)}-nh
    &\ge
      \mean\norm{S}_+
      -
      H(S|T)
      \ge
      \mean\norm{S\cap T}_-.
  \end{align}
  Maximizing the right-hand side over $\tau$ yields the claim.
\end{proof}

Theorems \ref{theoMISecondary}, \ref{theoMaxRepSecondary}, and
\ref{theoRedundancySecondary} can be wrapped up as a proposition that
compares power-law rates of various sublinear sequences, collectively
called Hilberg exponents \cite{Debowski15d, Debowski21} after
Hilberg's law. Hilberg's law is a hypothetical power law pertaining to
mutual information of natural language \cite{Hilberg90}---recently
studied under the name of the neural scaling law
\cite{HestnessOther17, KaplanOther20,
  HenighanOther2020, HernandezOther21}. Let us consider the Shannon
mutual information
\begin{align}
  I(X;Y):=H(X)+H(Y)-H(X,Y)
\end{align}
and the algorithmic mutual information
\begin{align}
  J(u;v):=K(u)+K(v)-K(u,v).
\end{align}
To be precise, the Hilberg exponent of a sequence is defined as
\begin{align}
  \hilberg_{n\to\infty} F(n):=
  \kwad{\limsup_{n\to\infty}
  \frac{\log F(n)}{\log n}}_+
  .
\end{align}
We have $\hilberg_{n\to\infty} n^\beta=\beta$ if $\beta\ge 0$.  The
next two propositions somewhat extend the results stated in
\cite[Theorems 1--3]{Debowski11b} and \cite[Section
8.4]{Debowski21}. The first theorem applies identity (\ref{ExcessII}),
which was discovered after publishing \cite{Debowski11b}.
\begin{theorem}
  \label{theoMI}
  For a stationary process $(X_i)_{i\in\mathbb{Z}}$ over alphabet
  $\mathbb{X}$, we have
\begin{align}
  1&\ge \beta^{\mathcal{B}}_\psi:=
    \hilberg_{n\to\infty}
    \okra{\mean\norm{\Gamma^{\mathcal{B}}_\psi(X_1^n)}-nh}=
    \hilberg_{n\to\infty}
    \mean J^{\mathcal{B}}_\psi(X_1^n;X_{n+1}^{2n})
    \nonumber\\
  &\ge
    \beta^{\mathcal{F}}_\psi:=
    \hilberg_{n\to\infty}
    \okra{\mean\norm{\Gamma^{\mathcal{F}}_\psi(X_1^n)}-nh}=
    \hilberg_{n\to\infty}
    \mean J^{\mathcal{F}}_\psi(X_1^n;X_{n+1}^{2n})
    \nonumber\\
   &\ge
    \beta^{\mathcal{G}}_\psi:=
    \hilberg_{n\to\infty}
    \okra{\mean\norm{\Gamma^{\mathcal{G}}_\psi(X_1^n)}-nh}=
    \hilberg_{n\to\infty}
    \mean J^{\mathcal{G}}_\psi(X_1^n;X_{n+1}^{2n})
    \nonumber\\
   &\ge
    \beta_K:=
    \hilberg_{n\to\infty}
    \okra{\mean K(X_1^n)-nh}=
    \hilberg_{n\to\infty}
    \mean J(X_1^n;X_{n+1}^{2n})
    \nonumber\\
  &\ge
    \beta_H:=
    \hilberg_{n\to\infty}
    \okra{H(X_1^n)-nh}=
    \hilberg_{n\to\infty}
    I(X_1^n;X_{n+1}^{2n})
    \ge 0.
\end{align}
\end{theorem}
\begin{proof}
  The claim follows by inequalities
  \begin{align}
    \mean\norm{\Gamma^{\mathcal{B}}_\psi(X_1^n)}
    \ge
    \mean\norm{\Gamma^{\mathcal{F}}_\psi(X_1^n)}
    \ge
    \mean\norm{\Gamma^{\mathcal{G}}_\psi(X_1^n)}
    \ge
    \mean K(X_1^n)-c
    \ge
    H(X_1^n)-c
  \end{align}
  for some $c>0$ and by identity
  \begin{align}
  \label{ExcessII}
    \hilberg_{n\to\infty} \okra{F(n)-nf}
    =
    \hilberg_{n\to\infty} \okra{2F(n)-F(2n)},
  \end{align}
  which holds if $\lim_{n\to\infty} F(n)/n=f$ and $F(n)\ge nf$
  \cite[Theorem 1]{Debowski21b}.
\end{proof}
The Hilberg exponents $\beta_H$ and $\beta_K$ equal zero for
finite-state sources but they are strictly positive and can be
arbitrarily close to $1$ for Santa Fe processes, see \cite{Debowski06,
  Debowski09, Debowski11b, Debowski21, Debowski21b, Hutter21,
  Debowski23} for the corresponding mathematical
framework. Consequently, the Hilberg exponents of the minimal codes
are also positive in the latter case.

The next statement resumes
Theorems \ref{theoMISecondary}, \ref{theoMaxRepSecondary},
\ref{theoRedundancySecondary}, and \ref{theoMI}.
\begin{theorem}
  Let $\mathcal{J}\in\klam{\mathcal{G},\mathcal{F},\mathcal{B}}$.
  For a stationary process $(X_i)_{i\in\mathbb{Z}}$ over alphabet
  $\mathbb{X}$, we have
  \begin{align}
    \label{Sandwich}
  \hilberg_{n\to\infty}\sup_{\tau}
  \mean\norm{\Sigma^{\mathcal{J}}_\psi(X_1^n)\cap \tau(X_{-\infty}^{0})}
  &\le \beta^{\mathcal{J}}_\psi\le
  \hilberg_{n\to\infty}
    \mean\norm{\Sigma^{\mathcal{J}}_\psi(X_1^n)}.
  \end{align}
\end{theorem}
\begin{proof}
  The claim follows by Theorems \ref{theoMISecondary},
  \ref{theoMaxRepSecondary}, \ref{theoRedundancySecondary}, and
  \ref{theoMI}, combined with inequality
  $\card V(S)\le \norm{S}$ which asymptotically blurrs
  the distinction between $\norm{S}_\pm$ and $\norm{S}$ under the
  Hilberg exponent.
\end{proof}

Thus the Hilberg exponents of minimal codes are sandwiched by the
rates of their vocabularies and their intersections.  We suppose that
the size of a vocabulary is close to the size of the vocabulary
intersection, up to a multiplicative constant depending on the
corresponding Hilberg exponent. 
Our conjecture is supported by the following observation.
\begin{definition}[sublinear and excessive vocabularies]
  A vocabulary transform is a mapping from strings to vocabularies.  A
  vocabulary transform $\Sigma$ is called sublinear if for any
  sequence $(x_i)_{i\in\mathbb{Z}}$, we have
  $\norm{\Sigma(x_1^n)}\le cn$ and
  \begin{align}
    \lim_{n\to\infty}\frac{\norm{\Sigma(x_1^n)}}{n}=0.
  \end{align}
  A vocabulary transform $\Sigma$ is called excessive if we have
  \begin{align}
    \norm{\Sigma(uv)}\ge
    \norm{\Sigma(u)}+\norm{\Sigma(v)}-\norm{\Sigma(u)\cap\Sigma(v)}.
  \end{align}
\end{definition}
\emph{Remark:} Intuitively, for an excessive vocabulary transform, the
vocabulary of a string concatenation is larger than the set-theoretic
sum of the vocabularies of the concatenated strings:
\begin{align}
  \norm{\Sigma(uv)}
  &\ge
    \norm{\Sigma(u)}+\norm{\Sigma(v)}-\norm{\Sigma(u)\cap\Sigma(v)}
    \nonumber\\
  &=
  \norm{\Sigma(u)\setminus\Sigma(v)}+
  \norm{\Sigma(v)\setminus\Sigma(u)}+\norm{\Sigma(u)\cap\Sigma(v)}.
\end{align}
\begin{theorem}
  For a sublinear and excessive random vocabulary transform $\Sigma$
  and a stationary process $(X_i)_{i\in\mathbb{Z}}$, we have
  \begin{align}
  \hilberg_{n\to\infty}
  \mean\norm{\Sigma(X_1^n)\cap \Sigma(X_{n+1}^{2n})}
  &=
  \hilberg_{n\to\infty}
    \mean\norm{\Sigma(X_1^n)}.
  \end{align}
\end{theorem}
\begin{proof}
  Excessivity and stationarity yield
  \begin{align}
    \mean\norm{\Sigma(X_1^n)\cap \Sigma(X_{n+1}^{2n})}\ge
    2\mean\norm{\Sigma(X_1^n)}-\mean\norm{\Sigma(X_1^{2n})},
  \end{align}
  whereas by identity (\ref{ExcessII}), sublinearity and the dominated
  convergence, we obtain
  \begin{align}
    \hilberg_{n\to\infty}
    \okra{2\mean\norm{\Sigma(X_1^n)}-\mean\norm{\Sigma(X_1^{2n})}}
    =
    \hilberg_{n\to\infty} \mean\norm{\Sigma(X_1^n)}.
  \end{align}
  The claim is obtained by noticing
  $\mean\norm{\Sigma(X_1^n)\cap \Sigma(X_{n+1}^{2n})}\le
  \mean\norm{\Sigma(X_1^n)}$.
\end{proof}

It is an open problem whether vocabulary transforms
$\Sigma^{\mathcal{J}}_\psi$ for
$\mathcal{J}\in\klam{\mathcal{G},\mathcal{F},\mathcal{B}}$ are
sublinear and excessive. If they are then the Hilberg exponents of the
minimal codes are equal to the Hilberg exponents of the corresponding
vocabularies. Thus, sandwich bound (\ref{Sandwich}) would be tight.

\subsection{False power law}
\label{secFalsePowerLaw}

The results presented so far suggest that the vocabulary size of the
minimal straight-line grammar grows at a power-law rate dictated by the
algorithmic mutual information of the source. The real picture is
somewhat more complicated, however. We encounter a competing false
power law that we will describe in this section. This result is a
development of an idea presented in the past.  In \cite{Debowski07d},
we have first observed that the vocabulary size for the minimal
grammar taken with respect to the Yang-Kieffer length
(\ref{YangKieffer}) grows at a nearly square-root rate that is
dictated by the entropy rate rather than the mutual information. This
observation was also restated in \cite[Theorem 4]{Debowski07c}. Here,
we would like to present an analog of this proposition for the local
grammar encoder. Due to the increasing cost of defining non-terminals,
this analog is an approximate power-law bound of form
\begin{align}
  \card V(\Sigma^{\mathcal{G}}_\psi(u))\gtrsim \exp\okra{\sqrt{c\ln \abs{u}}}.
\end{align}
This lower bound is asymptotically negligible if we take the Hilberg
exponent but is resembles a power law growth locally.

Indeed, let us take $v(n):=\exp\okra{\sqrt{c\ln n}}$. Then
$\hilberg_{n\to\infty} v(n)=0$. On the other hand, we may approximate
for $n$ in the vicinity of $n_0$,
\begin{align}
  \frac{v(n)}{v(n_0)}
  &=\exp\okra{\sqrt{c\ln n_0+c\ln\frac{n}{n_0}}-\sqrt{c\ln n_0}}
    \nonumber\\
  &\approx \exp\okra{\frac{c}{2\sqrt{c\ln n_0}}\ln\frac{n}{n_0}}
    =\okra{\frac{n}{n_0}}^{\beta_0},
    \quad
  \beta_0:=\frac{c}{2\ln v(n_0)}.
\end{align}

Having made this remark, we produce the exact formula for the minimal
straight-line code.
\begin{theorem}
  We have
  \begin{align}
    \log \card V(\Sigma^{\mathcal{G}}_\psi(u))
    >
    \sqrt{
    c_1
    \log\frac{\abs{u}}{L(u)+1}}
    -2\log(\log \abs{u}+1)-2c_1-c_2
    .
  \end{align}
\end{theorem}
\emph{Remark:} Since
$\abs{u}\ge \card V(\Sigma^{\mathcal{G}}_\psi(u))$ and $c_1\ge\log m$,
we may also apply upper bound
$\norm{\card V(\Sigma^{\mathcal{G}}_\psi(u))}/c_1\le
\norm{\abs{u}}/\log m$ to derive from (\ref{Necessity}) a looser
piecewise power law bound
\begin{align}
  \card V(\Sigma^{\mathcal{G}}_\psi(u))&>
  \okra{\frac{\abs{u}}{L(u)+1}-\alpha(u)}^{1/\alpha(u)}-m,
  &
    \alpha(u)&:=\ceil{\frac{\norm{\abs{u}}}{\log m}}.
\end{align}
\begin{proof}
  Observe that
  $\abs{w}c_1\le\norm{\card V(\Sigma^{\mathcal{G}}_\psi(u))}$ for any
  substring $w$ that appears twice in the minimal straight-line
  primary rule. Otherwise, if
  $\abs{w}c_1>\norm{\card V(\Sigma^{\mathcal{G}}_\psi(u))}$ then
  we could shorten the grammar by introducing the secondary rule
  $\card V(\Sigma^{\mathcal{G}}_\psi(u))\mapsto w$ and shifting the
  primary rule to the next position. Subsequently, we note that the
  maximal number of distinct substrings of length $k$ that may appear
  in the primary rule is
  $(\card V(\Sigma^{\mathcal{G}}_\psi(u))+m)^k$. Moreover, the length
  of the primary rule is greater than $\abs{u}/(L(u)+1)$ since the
  extension of any symbol in the primary rule must be shorter than
  $L(u)+1$. In view of these three observations, we obtain inequality
  \begin{align}
    \label{Necessity}
    (\card V(\Sigma^{\mathcal{G}}_\psi(u))+m)^k+1>
    \frac{\abs{u}}{L(u)+1}-k+1
  \end{align}
  for $k:=\ceil{\norm{\card V(\Sigma^{\mathcal{G}}_\psi(u))}/c_1}$.
  Notice now that $\norm{r}\ge\log(r+m)$ for a proper code
  $\psi$.  As a result, we obtain
  $2^{c_1 k}\ge \card V(\Sigma^{\mathcal{G}}_\psi(u))+m$. Thus we may
  bound further
  \begin{align}
    2^{c_1 (k+1)^2}\ge 2^{c_1 k^2}+k
    >
    \frac{\abs{u}}{L(u)+1}
    ,
  \end{align}
  which yields
  \begin{align}
    k+1>
    \sqrt{
    \frac{1}{c_1}
    \log\frac{\abs{u}}{L(u)+1}}
    .
  \end{align}
  Observing that
  $\abs{u}\ge \card V(\Sigma^{\mathcal{G}}_\psi(u))$ and
  $\norm{r}\le \log r+2\log(\log r+1)+c_2$ yields
  \begin{align}
    c_1(k-1)
    \le
    \norm{\card V(\Sigma^{\mathcal{G}}_\psi(u))}
    &\le
      \log \card V(\Sigma^{\mathcal{G}}_\psi(u))
      +2\log(\log \abs{u}+1)+c_2
      .
  \end{align}
  Combining the above two inequalities yields the claim.
\end{proof}

\subsection{Optimal block length}
\label{secOptimalBlockLength}

Let us change the area of interest and consider the minimal block
code. In this section, we will study the common length of its
secondary rules. We will see that for the proper minimal block code,
the length of secondary rules either is extremely large or it grows
essentially logarithmically with the input length.  Let us denote the
common length of rules in the minimal block vocabulary
$\Sigma^{\mathcal{B}}_\psi(u)$ as
\begin{align}
  L^{\mathcal{B}}_\psi(u):=\frac{d(\Sigma^{\mathcal{B}}_\psi(u))}{c_1}-1.
\end{align}

We arrive at the following specific bound which assumes that the
maximal length of a repetition grows at rate slower than a power law.
The proof indicates that other scenarios are possible if the entropy
rate vanishes.
\begin{theorem}
  Consider a stationary ergodic process $(X_i)_{i\in\mathbb{Z}}$ over
  alphabet $\mathbb{X}$.  Suppose that
  \begin{align}
    \label{ConditionBlockLength}
    h&>0,
    &
    \hilberg_{n\to\infty} \card V(\Sigma^{\mathcal{B}}_\psi(X_1^n))&<1 \text{ a.s.},
    &
    \hilberg_{n\to\infty} L(X_1^n)&=0 \text{ a.s.}
  \end{align}
  In this case, we have
  \begin{align}
    \lim_{n\to\infty}
    \frac{L^{\mathcal{B}}_\psi(X_1^n)}{\log n}\le \frac{1}{h} \text{ a.s.}
  \end{align}
\end{theorem}
\begin{proof}
  For the proper minimal block code, we may write
\begin{align}
  \label{BlockCodeLength}
  \norm{\Gamma^{\mathcal{B}}_\psi(X_1^n)}=
  \okra{\frac{\abs{u}}{L^{\mathcal{B}}_\psi(X_1^n)}-c_4}\bar\psi+
  \okra{\card V(\Sigma^{\mathcal{B}}_\psi(X_1^n))+c_5}\okra{L^{\mathcal{B}}_\psi(X_1^n)+1}c_1,
\end{align}
where $c_4\in[0,1]$, $c_5\in[0,2]$, and
$\bar\psi\in\kwad{c_1, \norm{\card
    V(\Sigma^{\mathcal{B}}_\psi(X_1^n))}}$.  The first term on the
right hand side roughly estimates the length of the primary rule,
whereas the second term expresses the total length of secondary rules
plus the remainder of the primary rule.

Solving the quadratic equation (\ref{BlockCodeLength}) for
$L^{\mathcal{B}}_\psi(X_1^n)$ yields
\begin{align}
  L^{\mathcal{B}}_\psi(X_1^n)
  =
  \frac{B-A}{2A}\okra{1\pm\sqrt{1-\frac{4AC}{(B-A)^2}}}
  ,
\end{align}
where we define
\begin{align}
  A&:=\okra{\card V(\Sigma^{\mathcal{B}}_\psi(X_1^n))+c_5}c_1,
  &
  B&:=\norm{\Gamma^{\mathcal{B}}_\psi(X_1^n)}+c_4\bar\psi,
  &
  C&:=\abs{u}\bar\psi.
\end{align}
We also define
\begin{align}
  L_1&:=\frac{C}{B-A},
  &
  L_2&:=\frac{B-A}{A}.
\end{align}
Then we may write
\begin{align}
  L^{\mathcal{B}}_\psi(X_1^n)
  =
  \frac{L_2}{2}\okra{1\pm\sqrt{1-\frac{4L_1}{L_2}}}.
\end{align}
Observe now that
$1-\frac{x}{2}(1+x)\le \sqrt{1-x}\le 1-\frac{x}{2}$. Thus we
obtain
\begin{align}
  \label{BlockCodeRuleSandwich}
  L^{\mathcal{B}}_\psi(X_1^n)
  \in
  \kwad{L_1,L_1\okra{1+\frac{4L_1}{L_2}}}\cup
  \kwad{L_2-L_1\okra{1+\frac{4L_1}{L_2}},L_2-L_1}.
\end{align}

Assume now (\ref{ConditionBlockLength}). The proper minimal block
code is universal so
  \begin{align}
    \lim_{n\to\infty} \frac{1}{n}\norm{\Gamma^{\mathcal{B}}_\psi(X_1^n)}=h \text{ a.s.}
  \end{align}
  Moreover, we have
  \begin{align}
    \limsup_{n\to\infty} \frac{\norm{n}}{\log n}= 1.
  \end{align}
  Hence for sufficiently large $n$, we obtain
  \begin{align}
    L(X_1^n)<L_2-L_1\okra{1+\frac{4L_1}{L_2}}.
  \end{align}
  Thus by $L^{\mathcal{B}}_\psi(X_1^n)\le L(X_1^n)$, we obtain
  \begin{align}
    L^{\mathcal{B}}_\psi(X_1^n)\le L_1\okra{1+\frac{4L_1}{L_2}},
  \end{align}
  which can be rewritten as the desired claim.
\end{proof}

\section{Infinite vocabularies}
\label{secInfiniteVocabularies}

In the previous sections, we studied minimal grammar-based codes with
a variable finite input-dependent vocabulary. There is a related
avenue of investigation, namely, minimal codes with a fixed infinite
input-independent vocabulary. If this vocabulary is known in advance
then it is sufficient to communicate only the primary part of the
grammar. It is interesting that there exist infinite input-independent
vocabularies which admit universal codes. These input-independent
vocabularies have a prohibitively high computational time complexity,
however.

In the following, we develop these ideas.  In Section
\ref{secVocabularyEntropies}, we investigate entropies of a stationary
process relative to fixed grammar vocabularies.  Section
\ref{secMeanField} presents a mean field theory of grammar-based
coding. Section \ref{secRankLists} discusses rank lists of universal
codes and some codes relative to rank lists.  For these developments,
we allow for infinite vocabularies. Mind that length $\norm{S}$ is
finite if and only if vocabulary $S$ is finite. We assume a proper
code $\psi$, as before.

\subsection{Vocabulary entropies}
\label{secVocabularyEntropies}

Let us investigate the asymptotic rate of the length of the primary
rule for a fixed but perhaps infinite vocabulary. This is quite a
natural development that rests on the sandwich bound
(\ref{MIPrimaryUpper})--(\ref{MIPrimaryLower}) and some pretty
standard toolbox of information theory.

We begin with the expected rate.
\begin{theorem}
  \label{theoPrimaryRate}
  Consider a stationary process $(X_i)_{i\in\mathbb{Z}}$ over
  alphabet $\mathbb{X}$. There exists limit
  \begin{align}
    h(S)
    :=
    \lim_{n\to\infty} \frac{\mean \norm{\Pi^{\mathcal{G}}_\psi(X_1^n|S)}}{n},
  \end{align}
  where
  $c_1 \le \mean\norm{\Pi^{\mathcal{G}}_\psi(X_1^n|S)} - nh(S)\le
  \norm{S}$.
\end{theorem}
\begin{proof}
  Define sequences
  $a_n^{(1)}:=\mean \norm{\Pi^{\mathcal{G}}_\psi(X_1^n|S)}-c_1$ and
  $a_n^{(2)}:=\norm{S}-\mean \norm{\Pi^{\mathcal{G}}_\psi(X_1^n|S)}$. By
  (\ref{MIPrimaryUpper})--(\ref{MIPrimaryLower}), they are
  subadditive, namely, $a_{n+m}^{(i)}\le a_n^{(i)}+a_m^{(i)}$. Hence
  by the Fekete lemma, there exists limit
  \begin{align}
    h(S)=\lim_{n\to\infty} \frac{a_n^{(1)}}{n}
    =\inf_{n\in\mathbb{N}} \frac{a_n^{(1)}}{n},
  \end{align}
  whereas for $\norm{S}<\infty$, we also have
  \begin{align}
    -h(S)=\lim_{n\to\infty} \frac{a_n^{(2)}}{n}
    =\inf_{n\in\mathbb{N}} \frac{a_n^{(2)}}{n}.
  \end{align}
  Thus
  $c_1 \le \mean\norm{\Pi^{\mathcal{G}}_\psi(X_1^n|S)} - nh(S)\le
  \norm{S}$.
\end{proof}

Entropies $h(S)$ have the infimum equal to the entropy rate.
\begin{theorem}
  \label{theoPrimaryRate}
  For a stationary process $(X_i)_{i\in\mathbb{Z}}$ over
  alphabet $\mathbb{X}$, we have
  \begin{align}
    \inf_S h(S)=h,
  \end{align}
  where $S$ ranges over all vocabularies.
\end{theorem}
\begin{proof}
  We observe that $\Pi^{\mathcal{G}}_\psi(X_1^n|S)$
  with a fixed $S$ is a prefix-free code for block $X_1^n$ so
  $h(S)\ge h$ by the source coding inequality.  Subsequently, applying
  the idea of block coding from the proofs of Theorems
  \ref{theoCriterionBlock} and \ref{theoUniversal}, we may bound
  \begin{align}
    &\inf_S \mean\norm{\Pi^{\mathcal{G}}_\psi(X_1^n|S)}
      \nonumber\\
    &\quad
      \le
    \min_{k\in\mathbb{N}}
    \kwad{
    \frac{n}{k}
    \okra{
    H(X_1^k)+2\log(\log m^k+1)+c_2
    }
    +(2k+1)c_1
    }
    .
  \end{align}
  Hence
  \begin{align}
    \inf_S h(S)
    &=
    \inf_S \inf_{n\in\mathbb{N}} \frac{\mean
    \norm{\Pi^{\mathcal{G}}_\psi(X_1^n|S)}}{n}
    =
    \inf_{n\in\mathbb{N}}
    \inf_S
    \frac{\mean
      \norm{\Pi^{\mathcal{G}}_\psi(X_1^n|S)}}{n}
      \nonumber\\
    &\le
    \inf_{k\in\mathbb{N}}
    \frac{
    H(X_1^k)+2\log(\log m^k+1)+c_2
    }{k}
    =
    h
    .
  \end{align}  
  Confronting $h(S)\ge h$ and
  $\inf_S h(S)\le h$, we infer $\inf_S h(S)=h$.
\end{proof}

The previous theorem has its almost sure counterpart.
\begin{theorem}
  \label{theoPrimaryRateAS}
  Consider a stationary ergodic process $(X_i)_{i\in\mathbb{Z}}$ over
  alphabet $\mathbb{X}$. We have
  \begin{align}
    \lim_{n\to\infty}
    \frac{\norm{\Pi^{\mathcal{G}}_\psi(X_1^n|S)}}{n}
    &=
    h(S)
    \text{ a.s.}
  \end{align}
\end{theorem}
\begin{proof}
  Define random variables
  $A_n^{(1)}:=\norm{\Pi^{\mathcal{G}}_\psi(X_1^n|S)}-c_1$. By
  (\ref{MIPrimaryLower}), they are subadditive in the sense of
  $A_{n+m}^{(1)}\le A_n^{(1)}+A_m^{(1)}\circ T^n$, where $T$ is the
  shift operation.  Hence Kingman's subadditive ergodic theorem
  \cite[Theorem 8.4]{Gray09} yields
  \begin{align}
    \lim_{n\to\infty}\frac{A^{(1)}}{n}
    =\lim_{n\to\infty} \frac{\mean A_n^{(1)}}{n}
    =\inf_{n\in\mathbb{N}} \frac{\mean A_n^{(1)}}{n}
    =h(S) \text{ a.s.}
  \end{align}
\end{proof}

\subsection{Mean field theory}
\label{secMeanField}

The theory of sublinear effects in the minimal grammar-based coding
rests on sandwich bounds (\ref{Sandwich}), which make the results less
elegant. Here we would like to advertise some simplified analysis that
arises when we try approximating expectation
$\mean
\norm{\Pi^{\mathcal{G}}_\psi(X_1^n|\Sigma^{\mathcal{G}}_\psi(X_1^n))}$
with quantity
$n\mean h(\Sigma^{\mathcal{G}}_\psi(X_1^n))+\mean
\norm{\Sigma^{\mathcal{G}}_\psi(X_1^n)}$ by an unjustified analogy to
the sandwich bound
$c_1\le\mean \norm{\Pi^{\mathcal{G}}_\psi(X_1^n|S)}-nh(S)\le \norm{S}$
for a fixed vocabulary $S$.

We will call such an approximation the mean field approximation.  Let
us put the mean field code length and the mean field vocabulary,
\begin{align}
  \label{ExpectedCodeI}
  B(n)&:=\min_{S}\kwad{nh(S)+2\norm{S}_+},
  \\
  \label{ExpectedCodeII}
  \Sigma(n)&:=\argmin_{S}\kwad{nh(S)+2\norm{S}_+}.
\end{align}
We have 
\begin{align}
  \mean \norm{\Gamma^{\mathcal{G}}_\psi(X_1^n)}
  &=
    \mean \min_{S} \kwad{\norm{\Pi^{\mathcal{G}}_\psi(X_1^n|S)}+\norm{S}_+}
    \nonumber\\
  &\le
  \min_{S}\kwad{\mean \norm{\Pi^{\mathcal{G}}_\psi(X_1^n|S)}+\norm{S}_+}
    \nonumber\\
  &\le
  B(n)
  \le
  \mean\kwad{
  n
  h(\Sigma^{\mathcal{G}}_\psi(X_1^n))
  +
  2\norm{\Sigma^{\mathcal{G}}_\psi(X_1^n)}_+
  }
  .
\end{align}

As we will see next, the power-law rate of the mean field vocabulary
$\norm{\Sigma(n)}$ is equal to the power-law rate of the mean field
mutual information $2B(n)-B(2n)$ and the mean field redundancy
$B(n)-nh$.  The following proposition is a solution to the more
abstract task stated in \cite[Chapter 8, Problem 5]{Debowski21}.
\begin{theorem}
  We have
  \begin{align}
    0
    \le
    2\norm{\Sigma(n)}_+
    &\le
    2B(n)-B(2n)
    \le
    2\norm{\Sigma(2n)}_+,
    \\
    \lim_{n\to\infty} \frac{B(n)}{n}
    &=
    h,
    \\
    \lim_{n\to\infty} \frac{\norm{\Sigma(n)}_+}{n}
    &=
    0,
    \\
    \lim_{n\to\infty} h(\Sigma(n))
    &=
    h,
    \\
    \hilberg_{n\to\infty} [B(n)-nh]
    &=
    \hilberg_{n\to\infty} [2B(n)-B(2n)]
    =
      \hilberg_{n\to\infty} \norm{\Sigma(n)}.
  \end{align}
\end{theorem}
\emph{Remark:}
In consequence, we obtain inequality
\begin{align}
  \hilberg_{n\to\infty} \kwad{
  \mean
  \norm{\Gamma^{\mathcal{G}}_\psi(X_1^n)}
  -
  nh
  }
  &\le
  \hilberg_{n\to\infty} \norm{\Sigma(n)}
  .
\end{align}
The mean field vocabulary growth upper bounds the true redundancy
rate.
\begin{proof}
  The first claim follows by
  \begin{align}
    2B(n)-B(2n)
    &\le
      2nh(\Sigma(2n))+4\norm{\Sigma(2n)}_+
      \nonumber\\
    &\quad -2nh(\Sigma(2n))-2\norm{\Sigma(2n)}_+=
      2\norm{\Sigma(2n)}_+.
  \end{align}
  Analogously,
  \begin{align}
    2B(n)-B(2n)
    &\ge
      2nh(\Sigma(n))+4\norm{\Sigma(n)}_+
    \nonumber\\
    &\quad
      -2nh(\Sigma(n))-2\norm{\Sigma(n)}_+=
      2\norm{\Sigma(n)}_+.
  \end{align}  
  The second claim follows by $\inf_{S} h(S)=h$ and
  \begin{align}
    \inf_{S} h(S)
    \le
    \lim_{n\to\infty} \frac{B(n)}{n}
    \le
    \inf_{S}
    \lim_{n\to\infty} \frac{nh(S)+2\norm{S}_+}{n}
    =
    \inf_{S} h(S)
    .
  \end{align}
  Hence we obtain
  \begin{align}
    \lim_{n\to\infty} \frac{\norm{\Sigma(n)}_+}{n}
    =
    \lim_{n\to\infty} \frac{2B(n)-B(2n)}{n}
    =
    0
  \end{align}
  and also
  \begin{align}
    \lim_{n\to\infty} h(\Sigma(n))
    =
    \lim_{n\to\infty} \frac{B(n)-2\norm{\Sigma(n)}_+}{n}
    =
    h.
  \end{align}
  The last claim follows from the first claim by $B(n)\ge nh$ and
  identity (\ref{ExcessII}).
\end{proof}

\subsection{Rank lists}
\label{secRankLists}

In this section, we will present some development that rests on the
idea of rank lists for incomplete distributions and the harmonic bound
proved in Lemma \ref{theoHarmonicBound}. We are not aware of a
similar construction in the literature. 

In the preceding exposition, we mentioned that we allowed infinite
vocabularies.  Let us make this definition---its statement is
analogous to the developements of Section \ref{secUniversality}.
\begin{definition}[adapted and universal vocabularies]
  A vocabulary $S$ is called adapted to a stationary ergodic process
  $(X_i)_{i\in\mathbb{Z}}$ over alphabet $\mathbb{X}$ if
  $h(S)=h$. A vocabulary $S$ is called universal if it is adapted to
  any stationary ergodic process $(X_i)_{i\in\mathbb{Z}}$ over
  alphabet $\mathbb{X}$.
\end{definition}

It is may seem surprising that universal vocabularies exist. Yet,
there are as many of them as there are distinct universal codes.  To
show it, we will first consider a few further definitions.  For a
vocabulary $S$, we define the rank of string $u$ as
\begin{align}
 R(u|S):=\inf\klam{\infty}\cup\klam{r\ge 1: S(r)=u}. 
\end{align}
Now two hair-splitting concepts, which will be useful later.
\begin{definition}[admissible vocabulary] 
  A flat vocabulary $S$ is called admissible for an incomplete
  distribution $Q$ if $R(u|S)\le 1/Q(u)$ for every string $u$ such
  that $Q(u)>0$.
\end{definition}
\begin{definition}[rank list]
  A flat vocabulary $S$ is called a rank list of an incomplete
  distribution $Q$ if $Q(S(r))\ge Q(S(s))$ for all $r\le s$.
\end{definition}
\begin{theorem}
  Any rank list of an incomplete distribution $Q$ is admissible for
  that distribution.
\end{theorem}
\begin{proof}
  Let $S$ be a rank list of $Q$. By the harmonic bound from Lemma
  \ref{theoHarmonicBound}, we obtain $R(u|S)\le 1/Q(u)$.
\end{proof}

Computing rank lists may have a prohibitive time complexity but it
leads to some interesting theoretical considerations. Here we define
two important cases of rank lists: First, the rank list $S_B$ of a
prefix-free code $B$ is the rank list of incomplete distribution
$Q_B(u)=2^{-\abs{B(u)}}$. Second, the Kolmogorov rank list $S_K$ is
the rank list of incomplete distribution $Q_K(u)=2^{-K(u)}$.

Now we name the construction that we studied in Section
\ref{secVocabularyEntropies}.
\begin{definition}[wrapped code]
  Let $S$ be a vocabulary. The wrapped code with respect to vocabulary
  $S$ is defined as function $u\mapsto\psi^*(\Pi^{\mathcal{G}}_\psi(u|S))$.
\end{definition}
We call this code wrapped since it uses strings from vocabulary $S$ as
a kind of black boxes referred by pointers determined by the ranking
of $S$.

By means of the wrapped code, we can show that the rank list of a
universal code is a universal vocabulary.  In the following, we still
tacitly assume proper coding.
\begin{theorem}
  \label{theoWrapped}
  Let $Q$ be an incomplete distribution and let $S$ be a vocabulary
  admissible for $Q$.  The wrapped code with respect to $S$ satisfies
  \begin{align}
    \label{WrappedBound}
    \norm{\Pi^{\mathcal{G}}_\psi(u|S)}
    &\le \norm{1/Q(u)}.
  \end{align}
  Consequently, if incomplete distribution $Q$ is adapted or universal
  then so are the wrapped code and the vocabulary $S$.
\end{theorem}
\emph{Remark:} The length of the wrapped code is subadditive by
inequalities (\ref{MIPrimaryUpper})--(\ref{MIPrimaryLower}). We may
treat this construction as a relatively low-cost transformation of an
arbitrary universal incomplete distribution into a related universal
code which is subadditive.  We note that the wrapped code word can be
shorter than the upper bound (\ref{WrappedBound}) if the corresponding
$Q$-probability is smaller than dictated by the harmonic bound.
\begin{proof}
  By the admissibility of $S$, we have $R(u|S)\le 1/Q(u)$.  Now we
  observe a simple upper bound
  $\norm{\Pi^{\mathcal{G}}_\psi(u|S)}\le \norm{R(u|S)}$ since we
  can abstract the whole string $u$ as a single non-terminal from $S$.
  Hence, we obtain (\ref{WrappedBound}).

  Code $u\mapsto\psi^*(\Pi^{\mathcal{G}}_\psi(u|S))$ is prefix-free
  since vocabulary $S$ is fixed. Its adaptedness or universality
  follows by adaptedness or universality of $Q$, inequality
  (\ref{WrappedBound}), and the observation that for any sequence
  $(x_i)_{i\in\mathbb{Z}}$ we have
  \begin{align}
    \lim_{n\to\infty}\frac{
    \norm{1/Q(x_1^n)}
    }{\log 1/Q(x_1^n)}
    =1
  \end{align}
  since $\lim_{n\to\infty} Q(x_1^n)=0$.
  
  Consider the fixed or an arbitrary stationary ergodic process
  $(X_i)_{i\in\mathbb{Z}}$ over alphabet $\mathbb{X}$. By the just
  established adaptedness or universality of code
  $u\mapsto\psi^*(\Pi^{\mathcal{G}}_\psi(u|S))$, we obtain that
  vocabulary $S$ is also adapted or universal since
  \begin{align}
    h(S)
    =
    \lim_{n\to\infty}
    \frac{\mean\norm{\Pi^{\mathcal{G}}_\psi(X_1^n|S)}}{n}
    =
    h
    .
  \end{align}  
\end{proof}

By the above construction, Zipf's law emerges for rank lists of
universal distributions. The result is as follows.
\begin{corollary}
  Let $Q$ be a universal distribution and let $S$ be a rank list of
  $Q$. For any stationary ergodic process $(X_i)_{i\in\mathbb{Z}}$
  over alphabet $\mathbb{X}$ with $h>0$, we have
  \begin{align}
    \lim_{n\to\infty}
    \frac{\log R(X_1^n|S)}{\log 1/Q(X_1^n)}=1
    \text{ a.s.}
  \end{align}
\end{corollary}
\emph{Remark:} It is interesting what happens for $h=0$ or
non-stationary processes.
\begin{proof}
  By the assumptions, we have
  $\norm{\Pi^{\mathcal{G}}_\psi(u|S)}\le \norm{R(u|S)}\le
  \norm{1/Q(u)}$.  By Theorem \ref{theoWrapped}, the wrapped code is
  universal. Hence $\norm{R(X_1^n|S)}/n$ is asymptotically sandwiched
  between $h$ and $\norm{1/Q(X_1^n)}/n$. The last quantity tends to
  $h$ for any stationary ergodic process
  $(X_i)_{i\in\mathbb{Z}}$. Thus, if $h>0$, then
  \begin{align}
    \lim_{n\to\infty}
    \frac{\log R(X_1^n|S)}{\log 1/Q(X_1^n)}
    =
    \lim_{n\to\infty}
    \frac{\norm{R(X_1^n|S)}}{\norm{1/Q(X_1^n)}}
    =
    1
    \text{ a.s.}
  \end{align}
\end{proof}

The above results have an intriguing philosophical interpretation. We
may consider the problem of universal coding as a certain formal model
of learning from empirical data. The particular data are stationary
ergodic processes and the learner is expected to learn to predict them
optimally.  Since the optimal prediction is implied by the optimal
compression \cite{DebowskiSteifer22, Debowski23b}, it is sufficient
that the learner applies a certain universal code. It may seem that
universal coding must consist in acquiring a certain specific
knowledge from the data that helps to compress only these particular
data. However, the existence of universal vocabularies challenges this
view.

To compress any data optimally, it is sufficient to know and apply a
source-independent rank list. Speaking metaphorically, so as to
predict the real world optimally, instead of learning pecularities of
this world, it is sufficient to learn absolute patterns of
mathematics. Of course, this is less suprising if we admit that the
atlas of mathematical patterns may contain pecularities of all
possible worlds. Moreover, this result seems of little practical
importance because it is much faster to learn pecularities of this
world from what we experience than to look them up in the atlas of
mathematical patterns.  With its power and perils, the situation may
resemble the framework of Kolmogorov complexity \cite{Solomonoff64,
  Kolmogorov65en2, Chaitin75, LiVitanyi08}.  Yet, it is still somewhat
baffling that we can catalog all possible patterns in a way that
satisfies Zipf's law and enables universal data compression.

\section{Finite vocabularies}
\label{secFiniteVocabularies}

In this section, we will scale down the concept of universal infinite
vocabularies to finite empirically driven rank lists that also define
universal codes. In Section \ref{secTrimmedCodes}, we inspect what
happens when we restrict an infinite universal vocabulary to a finite
prefix. In Section \ref{secEmpiricalCodes}, we apply a so restricted
code to the empirical rank list. We show that this yields a universal
code, as well.

\subsection{Trimmed codes}
\label{secTrimmedCodes}

Formally, the wrapped code is not a grammar-based code because its
code words lack the definition of the applied rank list. If we need an
exact grammar-based code then we may minimize grammars with respect to
the restrictions of a fixed rank list. We define the restriction
$S\upharpoonright k$ of a vocabulary $S$ through
$V(S\upharpoonright k):=k$ and $(S\upharpoonright k)(r):=S(r)$ for
$1\le r\le k$.
\begin{definition}[trimmed code]
  The trimmed grammar transform $\Gamma_\psi^S(u)$ with respect to an
  infinite vocabulary $S$ is defined as the straight-line grammar
  \begin{align}
    G=(S\upharpoonright k)\oplus \Pi^{\mathcal{G}}_\psi(u|S\upharpoonright k)
  \end{align}
  with the smallest $k$ that minimizes length $\norm{G}$. By an
  analogy to the prior notations, vocabulary $S\upharpoonright k$ for
  this particular $k$ is denoted $\Sigma_\psi^S(u)$.  The trimmed code
  is denoted
  \begin{align}
    B_\psi^S(u):=\psi^*(\Gamma_\psi^S(u)).
  \end{align}
\end{definition}

It is quite obvious that the trimmed code is universal for a universal
rank list. First, we need a simple lemma, however.
\begin{theorem}
  \label{theoRestrictionEntropy}
  We have $\inf_{k\in\mathbb{N}} h(S\upharpoonright k)=h(S)$ for any
  vocabulary $S$.
\end{theorem}
\begin{proof}
  We observe that for each string $u$ there exists a certain number
  $k(u)$ such that
  $\Pi^{\mathcal{G}}_\psi(u|S\upharpoonright
  k(u))=\Pi^{\mathcal{G}}_\psi(u|S)$ because the shortest primary rule
  applies only a finite number of distinct non-terminals. Taking the
  maximum $k(n):=\max_{u\in\mathbb{X}^n} k(u)$, we obtain
  $\Pi^{\mathcal{G}}_\psi(X_1^n|S\upharpoonright
  k(n))=\Pi^{\mathcal{G}}_\psi(X_1^n|S)$ almost surely. Hence for the
  fixed stationary process $(X_i)_{i\in\mathbb{Z}}$ over alphabet
  $\mathbb{X}$, we may write
  \begin{align}
    \inf_{k\in\mathbb{N}} h(S\upharpoonright k)
    &=
      \inf_{k\in\mathbb{N}}
      \inf_{n\in\mathbb{N}} \frac{\mean
      \norm{\Pi^{\mathcal{G}}_\psi(X_1^n|S\upharpoonright k)}}{n}
      =
      \inf_{n\in\mathbb{N}}
      \inf_{k\in\mathbb{N}}
      \frac{\mean
      \norm{\Pi^{\mathcal{G}}_\psi(X_1^n|S\upharpoonright k)}}{n}
      \nonumber\\
    &\le
      \inf_{n\in\mathbb{N}}
      \frac{\mean
      \norm{\Pi^{\mathcal{G}}_\psi(X_1^n|S\upharpoonright k(n))}}{n}
      =
      \inf_{n\in\mathbb{N}}
      \frac{\mean
      \norm{\Pi^{\mathcal{G}}_\psi(X_1^n|S)}}{n}
      =
      h(S)
    .
  \end{align}
  On the other hand,
  $\inf_{k\in\mathbb{N}} h(S\upharpoonright k)\ge h(S)$ since
  $\norm{\Pi^{\mathcal{G}}_\psi(u|S)}\le
  \norm{\Pi^{\mathcal{G}}_\psi(u|S\upharpoonright k)}$.
\end{proof}

Now we can demonstrate the universality.
\begin{theorem}
  \label{theoTrimmedUniversal}
  The trimmed code with respect to a vocabulary $S$ is
  adapted or universal if the vocabulary $S$ is adapted or universal.
\end{theorem}
\begin{proof}
  Consider the fixed or an arbitrary stationary ergodic source
  $(X_i)_{i\in\mathbb{Z}}$ over alphabet $\mathbb{X}$. By Theorem
  \ref{theoPrimaryRateAS}, we have
  \begin{align}
    \limsup_{n\to\infty}
    \frac{\norm{\Gamma_\psi^S(X_1^n)}}{n}
    &\le
    \inf_{k\in\mathbb{N}}
    \limsup_{n\to\infty}
      \frac{\norm{\Pi^{\mathcal{G}}_\psi(X_1^n|S\upharpoonright k)}}{n}
      \nonumber\\
    &=
    \inf_{k\in\mathbb{N}}
    h(S\upharpoonright k)
    =
    h(S)
    =
    h
    \text{ a.s.}
  \end{align}
  We also have the uniform bound
  $\norm{\Gamma_\psi^S(X_1^n)}\le c_1(n+2)$.  Hence code $B_\psi^S$ is
  adapted or universal by Barron's inequality \cite[Theorem
  3.1]{Barron85b} and the dominated convergence.
\end{proof}

Obviously, the trimmed code is significantly worse than the minimal
straight-line code or the wrapped code but it has a certain theoretical
aesthetic advantage: Its rates of mutual information and redundancy
are equal to the rate of its vocabulary.

Let us denote the pointwise mutual information
\begin{align} 
  J_\psi^S(u;v)
  &:=\norm{\Gamma_\psi^S(u)}+\norm{\Gamma_\psi^S(v)}-\norm{\Gamma_\psi^S(uv)}.
\end{align}
Here comes the advertised statement.
\begin{theorem}
  \label{theoMITrimmed}
  Suppose that vocabulary $S$ is adapted or universal.  Then
  for the fixed or an arbitrary stationary process
  $(X_i)_{i\in\mathbb{Z}}$ over alphabet $\mathbb{X}$, we have
\begin{align}
  &\hilberg_{n\to\infty}
    \okra{\mean\norm{\Gamma_\psi^S(X_1^n)}-nh}=
    \hilberg_{n\to\infty}
    \mean J_\psi^S(X_1^n;X_{n+1}^{2n})=
    \hilberg_{n\to\infty}
    \mean\norm{\Sigma_\psi^S(X_1^n)}
   \nonumber\\
  &\ge
    \hilberg_{n\to\infty}
    \okra{\mean\norm{\Gamma^{\mathcal{G}}_\psi(X_1^n)}-nh},
     \hilberg_{n\to\infty}
    \okra{\mean\norm{\Pi^{\mathcal{G}}_\psi(X_1^n|S)}-nh}.
\end{align}
\end{theorem}
\begin{proof}
  We apply the previously developed ideas with a twist.  First, we
  have
  \begin{align}
    J_\psi^S(u;v)
    \le 2\norm{\Sigma_\psi^S(uv)}_+
  \end{align}
  by an analogy to Theorem \ref{theoMISecondary}. On the other hand,
  the redundancy is lower bounded via inequality
  \begin{align}
    \mean\norm{\Gamma_\psi^S(X_1^n)}
    &=
    \mean\norm{\Pi^{\mathcal{G}}_\psi(X_1^n|\Sigma_\psi^S(X_1^n))}
    +
      \mean\norm{\Sigma_\psi^S(X_1^n)}_+
      \nonumber\\
    &\ge
    \mean\norm{\Pi^{\mathcal{G}}_\psi(X_1^n|S)}
    +
    \mean\norm{\Sigma_\psi^S(X_1^n)}_+
    \ge
    nh
    +
    \mean\norm{\Sigma_\psi^S(X_1^n)}_+
  \end{align}
  because
  $\norm{\Pi^{\mathcal{G}}_\psi(u|S)}\le
  \norm{\Pi^{\mathcal{G}}_\psi(u|S\upharpoonright k)}$. Obviously, we
  also have
  $\norm{\Gamma_\psi^S(u)}\ge\norm{\Gamma^{\mathcal{G}}_\psi(u)},
  \norm{\Pi^{\mathcal{G}}_\psi(u|S)}$.  Hence the claim follows by
  equality (\ref{ExcessII}).
\end{proof}

\subsection{Empirical codes}
\label{secEmpiricalCodes}

In this section, we consider trimmed codes applied to some particular
rank lists.  Let us consider the following two incomplete
distributions:
\begin{itemize}
\item The theoretical distribution for a stationary process
  $(X_i)_{i\in\mathbb{Z}}$ is
  \begin{align}
    Q_T(u)&:=w_{\abs{u}}P(X_1^{\abs{u}}=u),
    &
      w_k&:=\frac{1}{k+1}-\frac{1}{k+2}.
           \label{Theoretical}
  \end{align}
  Distribution $Q_T$ is adapted to $(X_i)_{i\in\mathbb{Z}}$ by the
  Shannon-McMillan-Breiman theorem
  \cite{Shannon48,Breiman57,Chung61,Barron85,AlgoetCover88}.  
\item For a threshold $p\in\mathbb{N}$, the discounted empirical
  distribution for a string $x_1^n$ is
  \begin{align}
    Q_p(u|x_1^n)&:=w_{\abs{u}}\max\klam{0,
                  \frac{\sum_{i=0}^{n-\abs{u}}
                  \boole{x_{i+1}^{i+\abs{u}}=u}-p+1}{n-\abs{u}+1}},
  \end{align}
  where we use the same weights $w_k$ as in (\ref{Theoretical}).
  Distribution $Q_1(u|x_1^n)$ is the empirical distribution, whereas
  distribution $Q_2(u|x_1^n)$ is called the repetition distribution,
  as it is supported on the set of repetitions.
\end{itemize}

Let $S_T(\cdot|x_1^n)$ be the rank list of the theoretical
distribution $Q_T(\cdot|x_1^n)$ and let $S_p(\cdot|x_1^n)$ be the rank
list of the discounted empirical distribution $Q_p(\cdot|x_1^n)$.  We
observe that the trimmed code $u\mapsto B_\psi^{S_T}(u)$, which
applies the theoretical rank list, is not universal but is adapted to
the particular stationary ergodic source. By contrast, we will show
that the trimmed code $u\mapsto B_\psi^{S_p(\cdot|u)}(u)$, which
applies the discounted empirical rank list, is universal. This result
is so desired that we will introduce a shorthand notation for this
code.
\begin{definition}[empirical code]
  For a threshold $p\in\mathbb{N}$, the empirical grammar transform is
  defined as $\Gamma_\psi^p(u):=\Gamma_\psi^{S_p(\cdot|u)}(u)$.  The
  empirical code is denoted
  \begin{align}
    B_\psi^p(u):=\psi^*(\Gamma_\psi^p(u)).
  \end{align}
\end{definition}

The empirical code works in an intuitive way. First, we compute the
empirical rank list for the compressed string. Second, we look for the
optimal restriction of this rank list that yields the best
compression. The empirical code resembles somewhat the complicated
algorithm by de Marcken \cite{DeMarcken96}, which was one of the
original impulses that motivated the field of grammar-based
coding. Nowadays similar algorithms are used for text tokenization in
large language models, viz.\ the byte-pair encoding (BPE)
\cite{SenrichHaddowBirch16}. We suspected that de Marcken's involved
algorithm is universal and now we can show that a somewhat similar
code is, indeed.

We will demonstrate universality of the emprical code without
establishing the convergence of the empirical rank list to the
theoretical one. The latter convergence need not hold if probabilities
of some strings are equal.
\begin{theorem}
  \label{theoEmpiricalUniversal}
  The empirical code is universal for any threshold $p\in\mathbb{N}$.
\end{theorem}
\begin{proof}
  Consider an arbitrary stationary ergodic process
  $(X_i)_{i\in\mathbb{Z}}$.  Let $\mathcal{S}_k$ be the set of
  vocabularies that are admissible for the restricted distribution
  $Q_T/(1+Q_T)|_{\mathbb{X}^k}$. Let $S_n$ be the rank list of the
  empirical distribution $Q_p(\cdot|X_1^n)$.  For each string $u$ with
  $Q_T(u)>0$, we have
  \begin{align}
    \limsup_{n\to\infty} 
    R(u|S_n)
    \le
    \lim_{n\to\infty}
    \frac{1}{Q_p(u|X_1^n)}
    =
    \frac{1}{Q_T(u)}
    <
    \frac{1}{Q_T(u)}+1
    \text{ a.s.}
    \label{RankListConvergence}
  \end{align}
  by the harmonic bound and the Birkhoff ergodic theorem.  Hence for
  any $k\ge 1$, we have $S_n\in\mathcal{S}_k$ for sufficiently large
  $n$ almost surely.

  Let $S_{nk}$ be the restriction of vocabulary $S_n$ such that
  $S_{nk}(r)=S_n(r)$ if $S_n(r)\in\mathbb{X}^k$, $S_{nk}(r)=\lambda$
  if $S_n(r)\not\in\mathbb{X}^k$ but $S_n(s)\in\mathbb{X}^k$ for
  some $s\ge r$, and $S_{nk}(r)$ is undefined else.  If
  $S_n\in\mathcal{S}_k$ then we have a uniform bound
  \begin{align}
    \norm{S_{nk}}
    \le
    (k+1)c_1
    \okra{\max_{u\in\mathbb{X}^k: Q_T(u)>0} \frac{1}{Q_T(u)}+1}
    <
    \infty
    ,
  \end{align}
  since the ranks of $k$-blocks are uniformly bounded by the harmonic
  bound.  Since $S_n\in\mathcal{S}_k$ for sufficiently large $n$
  almost surely, we obtain
  \begin{align}
    \limsup_{n\to\infty}
    \frac{\norm{\Gamma_\psi^p(X_1^n)}}{n}
    &\le
    \limsup_{n\to\infty}
      \frac{\norm{\Pi^{\mathcal{G}}_\psi(X_1^n|S_{nk})}+\norm{S_{nk}}_+}{n}
      \nonumber\\
    &\le
    \limsup_{n\to\infty}
    \frac{\norm{\Pi^{\mathcal{G}}_\psi(X_1^n|S_{nk})}}{n}
    \text{ a.s.}
  \end{align}
  
  Assume $S_n\in\mathcal{S}_k$ further. Applying a reasoning similar
  to the derivation of (\ref{MIPrimaryLower}), we obtain
  \begin{align}
    \norm{\Pi^{\mathcal{G}}_\psi(X_1^n|S_{nk})}
    &\le
    \frac{1}{k}\sum_{i=0}^{n-k}
    \norm{\Pi^{\mathcal{G}}_\psi(X_{i+1}^{i+k}|S_{nk})}
      +2(k+1)c_1
      \nonumber\\
    &\le
    \frac{1}{k}\sum_{i=0}^{n-k}
    \norm{1/Q_T(X_{i+1}^{i+k})+1}
    +2(k+1)c_1
    .
    \label{BlockBoundI}
  \end{align}
  Applying the Birkhoff ergodic theorem to (\ref{BlockBoundI})
  yields
  \begin{align}
    \limsup_{n\to\infty}
    \frac{\norm{\Pi^{\mathcal{G}}_\psi(X_1^n|S_{nk})}}{n}
    \le \frac{\mean\norm{1/Q_T(X_1^k)+1}}{k}
    \text{ a.s.}
    \label{BlockBoundII}
  \end{align}
  The infimum of the right-hand side of (\ref{BlockBoundII}) over $k$
  equals the entropy rate $h$. We notice that $k$ is arbitrary in
  (\ref{BlockBoundII}). Hence
  \begin{align}
    \limsup_{n\to\infty}
    \frac{\norm{\Gamma_\psi^p(X_1^n)}}{n}\le h
    \text{ a.s.}
  \end{align}
  We also have the uniform bound
  $\norm{\Gamma_\psi^p(X_1^n)}\le c_1(n+2)$.  Hence
  the empirical code is universal by Barron's inequality \cite[Theorem
  3.1]{Barron85b} and the dominated convergence.
\end{proof}

Unfortunately, the empirical code is not guaranteed to satisfy bounds
resembling Theorem \ref{theoMITrimmed}. It is so since it applies a
different vocabulary for each string and it results from merely a
local minimization. Yet, we suppose that the empirical code may
outperform the minimal block code in applications where we can allow
for a greater time complexity.  Of course, the greatest challenge for
a wider applicability of the empirical code is to estimate the optimal
number of non-terminals without computing all consecutive rank list
restrictions. Is there a fast computable restriction of the empirical
rank list that also secures universality?

\section{Conclusion}
\label{secConclusion}

The results of this paper can be located between information theory
and theoretical quantitative linguistics, with potential applications
to machine learning. The question that has been driving our interests
presents as follows: Does there exist a consistent tokenization
procedure that would yield simultaneously a universal code and a
mutual information bound in terms of the vocabulary size?  If such a
tokenization existed, it would connect Hilberg's law, nowadays known
as the neural scaling law in large language models, with Zipf's and
Heaps' laws of quantitative linguistics.

The theoretical results of this paper uncover the complexity of this
question by sketching both positive and negative results that concern
some particular tokenization procedures given by local grammar-based
codes. Whereas the minimal local grammar-based coding need not
efficiently recover the linguistically meaningful tokenization, it
provides a framework where Hilberg's, Zipf's, and Heaps' laws interact
on purely mathematical grounds.

Linking these results with empirical data is not straightforward,
however, as there are some obstacles such as the false power law
described in Section \ref{secFalsePowerLaw}. For this reason, it may
risky to draw conclusions from an empirical study of the effectively
computable minimal block code. The vocabulary size of this code may
not differentiate natural language texts and artificial data that do
or do not satisfy Hilberg's law, such as Santa Fe processes and
finite-state processes.  We think that the theory needs to be refined
yet before it can be tested empirically meaningfully. We hope that we
have sketched promising research directions.

\bibliographystyle{IEEEtran}
\bibliography{0-journals-abbrv,0-publishers-abbrv,ai,ql,mine,tcs,books,nlp}

\appendix

\section{Universality criteria}
\label{secCriteria}

In this section, for completeness and comparison, we prove the
universality criteria stated in Theorems \ref{theoCriterionMarkov} and
\ref{theoCriterionBlock}. Let $Q$ be an incomplete distribution and
$(X_i)_{i\in\mathbb{Z}}$ be a stationary ergodic process. By Barron's
inequality \cite[Theorem 3.1]{Barron85b} and the
Shannon-McMillan-Breiman theorem
\cite{Shannon48,Breiman57,Chung61,Barron85,AlgoetCover88}, we obtain
the lower bound
\begin{align}
  \label{CodeAS}
  \liminf_{n\to\infty} \frac{\kwad{-\log Q(X_1^n)}}{n}
  \ge
  \lim_{n\to\infty} \frac{\kwad{-\log P(X_1^n)}}{n}
  =
  h \text{ a.s.}
\end{align}
The analogous source coding inequality lower bounds the expectation.

It remains to prove the upper bounds. The following proofs are
analogous and based on the well known instance of the Toeplitz lemma
\begin{align}
  \lim_{k\to\infty}\frac{H(X_1^k)}{k}=
  \lim_{k\to\infty} H(X_i|X_{i-k}^{i-1})=h.
\end{align}
\begin{proof*}{Theorem \ref{theoCriterionMarkov}}
  Let $(X_i)_{i\in\mathbb{Z}}$ be stationary ergodic. By
  (\ref{UniversalityCriterionMarkov}),
  \begin{align}
    -\log Q(X_1^n)\le C(n,k)-\log\prod_{i=k+1}^{n}P(X_i|X_{i-k}^{i-1}).
  \end{align}
  Consequently, the Birkhoff ergodic theorem
  \cite{Birkhoff32,Garsia65} yields
  \begin{align}
    \limsup_{n\to\infty} \frac{\kwad{-\log Q(X_1^n)-C(n,k)}}{n}
    &\le
    H(X_i|X_{i-k}^{i-1}) \text{ a.s.}
  \end{align}
  This holds for any $k\ge 1$, so taking $k\to\infty$ gives
  \begin{align}
    \limsup_{n\to\infty} \frac{\kwad{-\log Q(X_1^n)}}{n}
    \le
    \lim_{k\to\infty} H(X_i|X_{i-k}^{i-1})
    =
    h \text{ a.s.}
  \end{align}
  Combining this with (\ref{CodeAS}) and the analogously derived upper
  bound for expectation, we establish universality of distribution
  $Q$.  
\end{proof*}

\begin{proof*}{Theorem \ref{theoCriterionBlock}}
  Let $(X_i)_{i\in\mathbb{Z}}$ be stationary ergodic. By
  (\ref{UniversalityCriterionBlock}),
  \begin{align}
    -\log Q(X_1^n)\le C(n,k)-\frac{1}{k}\log \prod_{i=0}^{n-k} P(X_{i+1}^{i+k}).
  \end{align}
  Consequently, the Birkhoff ergodic theorem
  \cite{Birkhoff32,Garsia65} yields
  \begin{align}
    \limsup_{n\to\infty} \frac{\kwad{-\log Q(X_1^n)-C(n,k)}}{n}
    &\le
    \frac{H(X_1^k)}{k} \text{ a.s.}
  \end{align}
  This holds for any $k\ge 1$, so taking $k\to\infty$ gives
  \begin{align}
    \limsup_{n\to\infty} \frac{\kwad{-\log Q(X_1^n)}}{n}
    \le
    \lim_{k\to\infty}
    \frac{H(X_1^k)}{k}
    =
    h \text{ a.s.}
  \end{align}
  Combining this with (\ref{CodeAS}) and the analogously derived upper
  bound for expectation, we establish universality of distribution
  $Q$.
\end{proof*}

\end{document}